\documentclass[num-refs]{nbdt-article}

\usepackage{siunitx}
\usepackage{float}

\papertype{Original Article}
\paperfield{Journal Section}

\title{How do we generalize?}

\author[1\authfn{1}]{Jessica Elizabeth Taylor PhD}
\author[1,2\authfn{1}]{Aurelio Cortese PhD}
\author[3,4\authfn{0}]{Helen C Barron  PhD}
\author[5\authfn{0}]{Xiaochuan Pan PhD}
\author[6\authfn{0}]{Masamichi Sakagami PhD}
\author[7\authfn{1}]{Dagmar Zeithamova PhD}

\affil[1]{The Department of Decoded Neurofeedback, Computational Neuroscience Laboratories, Advanced Telecommunications Research Institute International, Kyoto, Japan}
\affil[2]{Institute of Cognitive Neuroscience, University College London, UK}
\affil[3]{Medical Research Council Brain Network Dynamics Unit, Nuffield Department of Clinical Neurosciences, University of Oxford, Oxford, UK}
\affil[4]{Wellcome Centre for Integrative Neuroimaging, University of Oxford, FMRIB, John Radcliffe Hospital, Oxford, UK}
\affil[5]{Institute for Cognitive Neurodynamics, East China University of Science and Technology, Shanghai, China}
\affil[6]{Brain Science Institute, Tamagawa University, Tokyo, Japan}
\affil[7]{Department of Psychology, University of Oregon, United States}

\corraddress{Jessica Elizabeth Taylor PhD, Aurelio Cortese PhD, or Dagmar Zeithamova PhD}
\corremail{jessie.elizabeth.taylor@gmail.com, cortese.aurelio@gmail.com, or dasa@uoregon.edu}

\fundinginfo{National Institutes of Health (USA), Grant Number: R01-NS112366; JST ERATO (Japan), Grant Number: JPMJER1801}

\runningauthor{Taylor et al.}

\begin{document}

\maketitle

\begin{abstract}
Humans and animals are able to generalize or transfer information from previous experience so that they can behave appropriately in novel situations. What mechanisms--computations, representations, and neural systems--give rise to this remarkable ability? The members of this Generative Adversarial Collaboration (GAC) come from a range of academic backgrounds but are all interested in uncovering the mechanisms of generalization. We started out this GAC with the aim of arbitrating between two alternative conceptual accounts: (1) generalization stems from integration of multiple experiences into summary representations that reflect generalized knowledge, and (2) generalization is computed on-the-fly using separately stored individual memories. Across the course of this collaboration, we found that--despite using different terminology and techniques, and although some of our specific papers may provide evidence one way or the other--we in fact largely agree that both of these broad accounts (as well as several others) are likely valid. We believe that future research and theoretical synthesis across multiple lines of research is necessary to help determine the degree to which different candidate generalization mechanisms may operate simultaneously, operate on different scales, or be employed under distinct conditions. Here, as the first step, we introduce some of these candidate mechanisms and we discuss the issues currently hindering better synthesis of generalization research. Finally, we introduce some of our own research questions that have arisen over the course of this GAC, that we believe would benefit from future collaborative efforts.

\end{abstract}

Consider a scenario where you are bitten by a stranger’s dog, who happens to be a big German Shepherd. After this experience, you might find yourself starting to feel anxious around other German Shepherds, and other large breeds of dog, although you remain comfortable with your neighbor’s chihuahua. This transfer of danger information from the dog that bit you to other similar dogs is a classic example of generalization. Now consider another scenario. One day, your new co-worker arrives at the office wearing a velvet dress, although most of your co-workers dress casually. This reminds you of your college friend who loved to wear velvet dresses. Remembering how your college friend kept inviting you to come along with her to opera, you find yourself wondering if your new co-worker is also particular to the opera. While this scenario is very different from the one above, many (but not all) researchers would also consider this scenario to fit under the term “generalization”. In this paper, we will discuss how the term “generalization” has been defined and operationalized in multiple ways, with the aim to provide a glimpse into the breadth of phenomena that can be referred to as generalization. 
					
Because of the importance of generalization for adaptive learning and behavior, it has been studied in a variety of disciplines ranging from decision-making, perception, psychology, and memory neuroscience, to machine learning, and artificial intelligence. However, knowledge about the mechanisms that support generalization has not always been integrated across disciplines. In this paper we broadly describe what has been studied under the umbrella term “generalization”, discuss several candidate generalization mechanisms, and highlight the main challenges in synthesizing existing research and resolving among different generalization views. We acknowledge that although this Generative Adversarial Collaboration (GAC) set out with the aim of arbitrating between two seemingly opposing generalization mechanisms, in the course of discussions within our group, we realized that none of us consider these to be mutually exclusive or exhaustive accounts of generalization. Instead, generalization may be an outcome of multiple mechanisms, operating under different conditions and/or at different scales. With this in mind, we ourselves will use the term “generalization” to refer to the phenomenon (or phenomena) through which we flexibly draw upon experience to respond appropriately in novel situations or to novel stimuli, irrespective of the underlying mechanism.  We hope that our paper will increase awareness of different views on generalization and postulated mechanisms, that it will facilitate cross-talk between different labs, and that it will lead to a better synthesis of generalization research in the future.

\section{Proposed mechanisms of generalization}
Multiple mechanisms have been proposed to explain how information from past experience is generalized to novel situations or stimuli. In this section we describe several of the most well-supported mechanisms. While we here describe each of these mechanisms separately, we will discuss later in the paper that these are not necessarily mutually exclusive.  			

\subsection{Complementary learning systems}
How can humans and other animals both recall specific details of individual experiences and generalize information across these different experiences? For example, a squirrel has to remember a specific location where they hid their nuts \cite{Jacobs1991-lj} and humans can remember specific idiosyncratic details of individual events. However, to successfully generalize information for appropriate use in novel situations, learned information should reflect stable features of the environment and discard idiosyncratic details of individual experiences. Theories of generalization have to therefore address how memory may provide a basis for generalization while also supporting recall of individual events. 
					
A biologically plausible framework first proposed to explain how these distinct memory functions may be accomplished in the brain is the complementary learning systems theory \cite{McClelland1995-an}. According to this view, complementary neural learning systems coexist in the brain, with a division of labor between a fast and a slow learning system. The fast learning system, ascribed to the hippocampus, learns to encode individual events. In contrast, connections between neurons elsewhere in the brain change slowly and incrementally to support the formation of generalized memory representations which reflect statistical regularities across experiences. Idiosyncratic details that do not repeat regularly do not influence these generalized representations. At least two flavors of the complementary learning systems theory exist. Both agree on the role of the hippocampus in rapid encoding of specific events but these emphasize different brain regions to support the slow statistical learning underlying generalization: the striatum \cite{Poldrack2003-uc} or the neocortex \cite{OReilly2000-ti}. These two brain regions may support different forms of generalization rather than underpin competing theories of generalization.	

\subsection{Memory integration}
Memory integration theory of generalization \cite{Eichenbaum1999-oa} builds on findings that generalization may be rapid and rely on the hippocampus, demonstrating that complementary learning systems theory does not provide a full picture of generalization. Memory integration can result when existing memories are reactivated during encoding of current events that overlap in content, such as when they happen in the same place or contain some of the same elements \cite{Zeithamova2012-ut}. The current and reactivated information is then integrated into a combined memory representation that links elements from the distinct events, supporting generalization across events. Such generalization based on integrated memories may come with trade-offs, such as a confusion between directly experienced and inferred information \cite{Shohamy2008-qd, Zeithamova2012-ut, Varga2019-bn}. The memory integration and complementary learning systems views are not necessarily at odds with each other. Both ascribe to the hippocampus rapid encoding and retrieval of episodic memories, although the memory integration view additionally highlights how these functions may result in integration of a small number of experiences in service of rapid generalization. The complementary learning systems view, on the other hand, accounts for slow, incremental learning and for findings that learning and generalization are possible even without an intact hippocampus \cite{Knowlton1993-xf, Squire1995-uy}.
					
Several studies have provided empirical evidence for the existence of integrated memories in the hippocampus and their use in generalization (see \cite{Schlichting2017-aq} and \cite{Zeithamova2020-sh}, for reviews). Many studies have also implicated the medial prefrontal cortex (mPFC) \cite{Barron2013-iy, Bowman2018-pd, Kumaran2009-cn, Schlichting2015-rj} and hippocampal-mPFC interactions in generalization and memory integration \cite{Zeithamova2012-ut, Frank2019-wx, Gerraty2014-ji, Van_Kesteren2010-du, Van_Kesteren2014-ym}. A debate remains whether the hippocampus directly integrates memories that are then represented by mPFC or whether the hippocampus primarily represents separate memories that are then integrated by the mPFC. 
					
At a higher level, memories can be integrated in structured ways, as was first proposed with the concept of cognitive maps \cite{Tolman1948-sn}. Place cells \cite{OKeefe1971-uy, Okeefe1978-va} and grid cells \cite{Hafting2005-jf} may constitute the central building blocks for these processes. While place cells are primarily found within the hippocampus, grid cells can be observed in entorhinal cortex together with mPFC \cite{Hok2005-yg, Doeller2010-qv, Jacobs2013-kb, Yartsev2013-dg}. Crucially, these neurons display the same patterns of activity when agents navigate classic Euclidean space and non-spatial cognitive tasks \cite{Constantinescu2016-ys, Bellmund2018-vg, Mok2019-zg}. Thus, place and grid cells may afford a general computational mechanism whose function extends to selectively organize knowledge (about space, but also experiences, concepts, etc) into internal predictive maps \cite{Stachenfeld2017-tj, Whittington2020-uq}. This encapsulates the relationships between task-relevant variables \cite{Doeller2010-qv, Constantinescu2016-ys, Carpenter2015-vj, Bellmund2016-yz} , even at the level of very abstract structures such as words in language \cite{Vigano2021-gz} and social dimensions \cite{Park2020-ng}. In sum, the predictive cognitive map could arise as the consequence of memory integration, to support flexible generalization. This may allow, for example, an abstract map experienced in one sensory domain to be generalized to another sensory domain, or estimation of new trajectories using parameters learned in the past.

\subsection{On-the-fly generalization through co-activation of separate memories} While proponents of memory integration theory view generalization as something that results from the formation of a unique type of representation (integrated memories), others argue that both memory for individual experiences and generalization can result from the same type of representation (non-integrated memories) \cite{Curtis2019-ww, Hintzman1980-ms, Kinder2001-cb, Zaki2003-oq}. The REMERGE model (Recurrency and Episodic MEmory Results in GEneralization) offers an attractive explanation for how generalization may arise on-the-fly from a neural system that is optimized to orthogonalise memories \cite{Kumaran2012-gs}. Specifically, memories of individual events are stored as separate representations in the hippocampus. Generalization happens on-the-fly, at the time of retrieval in response to task demands, by co-activated or sequential activation of multiple individual memories. 

In the most extreme scenario, on-the-fly generalization can be viewed as a process of working with existing memory representations that themselves remain intact. However, there is also evidence that the process of on-the-fly computation, despite involving initially separate representations, may ultimately change the representations themselves so that they become integrated \cite{Carpenter2017-bi}. Notably, even though memories are initially stored separately, they are recirculated through the anatomical architecture of the hippocampal loop \cite{Koster2018-sb}. This allows the hippocampal output on one iteration to provide new input for the next iteration, thereby allowing integration of information across multiple episodes. In other words, once generalization is computed, it is sent back to the hippocampus and becomes a new, generalized memory construed from information that has been integrated across once-separate memories. This model maps onto the known neural circuitry of the hippocampus and neocortex, at the level of both anatomical and physiological connections. For example, big-loop recurrence can be implemented via the hippocampal architecture together with the laminar connectivity between the deep and superficial layers of the entorhinal cortex \cite{Koster2018-sb}n offers a major advantage over the more standard memory integration theory, as the system does not have to anticipate which memories would be most relevant to integrate during encoding. Instead, links between events can be dynamically computed on-the-fly using separate memories.	

\subsection{Offline generalization during sleep and rest}
The memory integration view often focuses on encoding processes. Specifically, its proponents highlight how reactivation of a past related event during the encoding of a new event can result in the formation of an integrated representation. By contrast, on-the-fly generalization assumes that memories of individual events are stored as separate representations that can be simultaneously reactivated when needed. We have already touched upon the possibility that memory integration and on-the-fly generalization may not be mutually exclusive (see also \cite{Zeithamova2020-sh}, for a review of evidence for each of these scenarios). Additionally, evidence from rodents \cite{Barron2020-bm, Wiltgen2007-jq} and humans \cite{Diekelmann2010-wy, Ellenbogen2007-se, Lau2011-hy, Schapiro2017-tw, Tompary2017-gd} indicates that memory representations can transform from specific to generalized/integrated spontaneously, during periods of rest or sleep. Recent work indicates a key role for sharp-wave ripples (SWR) in spontaneous generalization. SWRs occur in the hippocampus during periods of quiet wakefulness or sleep and are thought to reflect memory replay \cite{Buzsaki2015-qh, Joo2018-ho, Foster2017-xw}. They may extend beyond replay of directly experienced information to anticipate upcoming experience \cite{Dragoi2011-qp, Olafsdottir2015-wp}, reorder events according to a learned rule \cite{Liu2019-cu}, and/or join-the-dots between discrete but logically related experiences \cite{Barron2020-bm}. In this manner, hippocampal spiking activity during SWRs may provide an efficient means by which memories are integrated for generalization. Integration may happen especially when this is of biological advantage, as suggested by a study where hippocampal SWRs were uniquely found to reflect inferred relationships across events that resulted in reward \cite{Barron2020-bm}.
					
Coordinated activity between the hippocampus and neocortex during periods of sleep and rest is hypothesized to provide a potential signature for memory consolidation \cite{Joo2018-ho}, where hippocampal SWRs correlate with sleep spindles in the neocortex during slow-wave sleep \cite{Ji2007-dn, Qin1997-jl}. While this coupling between hippocampus and neocortex during sleep/rest may simply strengthen representations of learned information, it is plausible that hippocampal spiking activity also facilitates formation of generalized representations in the neocortex. Periods of rest and sleep may therefore provide an opportunity to integrate memories across neural circuits, to support flexible and adaptive behavior in the future.	

\subsection{Decision bound theory \& rule-based generalization}
Decision bound theory \cite{Ashby1992-pl, Ashby1993-bi, Maddox1993-oh} postulates that people can seamlessly generalize to new stimuli because they learn associations between a region of perceptual space and a response, rather than between specific stimuli and responses. The regions of perceptual space are divided by a ‘decision bound’ so that stimuli falling on one side of this bound are assigned one type of response and stimuli falling on the other side are assigned another. Importantly, a decision bound is best understood as a convenient descriptor of participants’ behavior rather than a postulated underlying mechanism, or something necessarily represented in the brain. 
					
In the Competition between Verbal and Implicit System (COVIS) model of categorization, Ashby and colleagues argue for distinct mechanisms of generalization depending on whether the decision bound is easily described verbally or not \cite{Ashby1998-uy, Gregory_Ashby_undated-rh}. The postulated implicit system of COVIS, attributed to the striatum, is thought to support stimulus-response mapping. It learns slowly and incrementally to associate motor responses with regions of perceptual space, akin to the slow-learning system postulated by the complementary learning systems theory. The decision bound can be inferred from behavioral responses but is not necessarily explicitly represented by the striatum or existing in people’s mind. By contrast, the postulated verbal system, attributed to the prefrontal cortex and later extended to include the hippocampus \cite{Nomura2006-av} derives explicit (verbalizable) decision bounds, or rules. The verbal system dominates when an easily verbalizable rule exists, typically when the optimal response is determined by a single stimulus feature (e.g., only color or only orientation).
				
Not all agree that there is a specific rule-based generalization mechanism, dissociable from a more automatic stimulus-response learning system \cite{Nosofsky2000-fl, Pothos2005-yt, Verguts2009-vw}. However, a number of studies have shown compelling dissociations between categorization based on verbalizable versus more implicit rules (for a review, see e.g., \cite{Maddox2004-kc}). For example, consider that a subject learns the association between short steep lines and one motor response (e.g., “A” button press), and long shallow lines and a different motor response (e.g., “B” button press). Now consider that the subject encounters new stimuli that fall on distinct sides of the same imaginary decision bound between the original A and B category, but in a previously untrained region of perceptual space. For example, the subject may encounter even shorter steeper lines and even longer shallower lines than before. None of the originally firing perceptual neurons are now firing for these lines because these are in a different region of perceptual space to that previously experienced (despite being separated by the same decision bound). This means that, if based on striatal-response mappings, then learning should not generalize to these new stimuli. Casale et al. (2012) \cite{Casale2012-dx} showed that this was indeed the case for a task putatively dominated by the implicit system (see also Smith et al., 2015 for a replication in macaques \cite{Smith2015-hz}, and Qadri et al., 2019 for a replication in pigeons \cite{Qadri2019-ib}). However, in a task putatively dominated by the verbal, rule-based system, Casale and colleagues found a successful generalization to a new region in perceptual space along the previously learned decision bound. Thus, rule-based generalization seems qualitatively unique and could account for the transfer of knowledge to new situations that are perceptually distinct from old situations (see also section 2.1.4).	

\section{Issues that are currently hindering better synthesis of generalization research}
Which theory of generalization best fits existing data? Despite the wealth of data on generalization, synthesis of existing studies to answer this question may be challenging. Many studies have been designed using  different definitions of generalization, and have been carried out in different neuroscience subfields, using a variety of tasks, research tools, brain regions and model species. These differences raise the following questions: Are we all investigating the same underlying generalization process? Are we investigating distinct types of generalization supported by different mechanisms? Or are we investigating distinct mechanisms--perhaps subserved by different brain regions--but with some fundamental underlying principles common to all of them? Some of the key challenges to the synthesis of existing data are discussed in this section.

\subsection{The same term for different things?}
\subsubsection{Confusing behavioral expression and underlying mechanism}
Once we consider the range of generalization theories discussed above, it becomes clear that an observable behavioral outcome (e.g., successful generalization judgment) should not be equated with an underlying mechanism (e.g., construction of a generalized representation).  Do successful generalization judgments arise because the participant formed a generalized knowledge representation during learning and now queries this knowledge? Do they arise because the participant integrated multiple experiences in response to task demands, transforming specific memories to generalized knowledge at the time of the generalization test \cite{Carpenter2017-bi}? Or does what appears as generalization behavior simply reflect encoding and retrieval of a series of specific memories, without the need for any generalized representations or any unique generalization mechanism \cite{Kumaran2012-gs}? 

At the mechanistic level, “generalization” could further involve, or refer to, the unique nature of an underlying representation, a transformation process, or an online computation that supports a generalization judgement. For example, memory integration theory emphasizes the nature of representations supporting generalization, where generalization behavior itself may be a simple recall of already-generalized representations. In the sleep literature, “generalization” would most commonly refer to the sleep-related formation of generalized representation or to the sleep-related transformation of initially specific memories to more generalized memories (with no immediate behavioral expression at the time of generalization). On-the-fly “generalization” refers to the retrieval and manipulation of multiple representations (something happening with representations). Yet, on-the-fly “generalization” may also involve transformation of representations, as discussed earlier. Different papers and different research subfields tend to focus on different aspects (online or offline computations, representations or transformation) of the postulated generalization mechanisms.
 
In this paper, we use the term “generalization” to refer to the observable behavioral outcome, irrespective of whether the underlying mechanism involved any generalization-unique representations or computations. Yet, even then, we are facing a breadth of phenomena that fall under the umbrella of generalization behaviors, as many real-world scenarios and laboratory tasks involve the transfer of past experience to novel situations. Whether or not they all share an underlying neural mechanism remains an open question. 

\subsubsection{“Similarity-based” versus “relational” versus “rule-based” generalization}
Similarity-based generalization occurs when information is transferred across multiple stimuli based on similarities in their features. This type of generalization is exemplified by Ivan Pavlov’s famous experiments in the early nineteen hundreds where dogs learned to predict the receipt of food from a bell sound. The conditioned response generalized such that the dogs started salivating not only in response to the original bell sound, but also to similar sounds. Most similarity-based generalization research focuses on perceptual similarity, while some studies in humans also consider semantic similarity \cite{Till2000-vz}. 

Relational generalization does not require similarities in stimulus features. Instead, relational generalization occurs when information is transferred across stimuli based on arbitrary learned associations. In a famous example of this, Brogden exposed dogs to repeated light-bell pairings before pairing the bell alone with a shock \cite{Brogden1939-kd}. This caused the bell to elicit a conditioned response to the shock. It also led to generalization of that conditioned response from the bell to the otherwise neutral (perceptually unrelated) light. Since the initial investigation of this type of relational generalization, coined sensory preconditioning \cite{Brogden1939-kd}, a number of studies have shown evidence for relational generalization across a range of different scenarios, including operant conditioning and stimulus-stimulus associations between neutral stimuli \cite{Yakovlev1998-qd}. 	

Rule-based generalization, described in models such as COVIS \cite{Ashby1998-uy, Gregory_Ashby_undated-rh}, is thought to rely on explicit verbalizable rules. A rule is typically based on a defining perceptual feature, such as choosing a response to a new stimulus solely based on whether it is red and ignoring all other perceptual features. As discussed in section 1.5., some have suggested that rule-based generalization is a special case of similarity-based generalization with attention selectively focused on a single dimension \cite{Nosofsky1986-pb}, while others argue that rule-based generalization is mechanistically unique \cite{Ashby1998-uy, Casale2012-dx, Rips1989-yd}. Furthermore, rule-based generalization can occur based on abstract, conceptual knowledge. For example, Shanks and Darby (1998) presented participants with various “foods” (labeled A, B, C, etc) either as individual items (e.g. A) and/or as paired combinations (e.g. AB) \cite{Shanks1998-xs}. Participants were asked to predict whether these foods were safe or predictive of an allergic response. Participants may have noticed an underlying rule: that foods led to opposite outcomes depending on whether they were presented individually or in combination. At test, in response to paired combinations whose components had only been trained as single items, some participants did not utilize this rule and instead used similarity-based generalization (e.g. predicting an allergic response to AB after learning that A causes allergic responses). Others were able to generalize using the underlying rule (e.g. predicting that AB is safe after learning that A causes allergic responses). While such abstract rule-based generalization seems especially distinct from similarity-based generalization, some argue it may still be supported by the same computational principles \cite{Verguts2009-vw}. 

\subsubsection{Rapid versus incremental learning-based generalization}
Historically, slow incremental learning has been highlighted by the complementary learning systems theory as the key mechanism of generalization, giving rise to memory representations that capture common themes across experiences \cite{McClelland1995-an}. However, rapid generalization can also be observed in behavior, and may have more adaptive significance.  For example, most people will rapidly learn to avoid all angry dogs after a bad experience with just one. Even rapid relational generalization is possible, with evidence showing that this can arise after each relationship has been experienced just once \cite{Zeithamova2010-ux}. The theories outlined earlier in the paper ascribe rapid generalization and incremental learning-based generalization to distinct neural systems, such as rapid generalization in the hippocampus in memory integration theory and slow generalization in the striatum and/or neocortex in complementary learning systems theory. Furthermore, rapid generalization is hard to achieve in computational models of incremental learning. This suggests that the underlying principles for rapid and incremental learning-based generalization may be different. Finally, we know that the rapid learning system may provide repetition and teaching signals to the slow-learning system, blurring the division of labor between rapid and incremental learning for generalization. For example, rapidly encoded hippocampal memories are thought to replay during sleep or quiet periods of rest, providing an opportunity to consolidate a more generalized version of those memories into the slow-learning cortex (see \cite{Lewis2011-ih, Witkowski2020-dh} for reviews).	

\subsubsection{Within-distribution versus out-of-distribution generalization}
Machine learning and predictive models reveal yet another way in which different types of generalization may manifest: as ‘within-distribution’ versus ‘out-of-distribution’ generalization. For within-distribution generalization, the data used to test an agent or model comes from the same sample distribution as that used to train it. For example, data from the original distribution can be deliberately held out from training such that it can be used subsequently for cross validation testing. On the other hand, for out-of-distribution generalization, a trained agent or model can be tested on data from a different distribution than the training data. For example, a model could be trained on medical imagery from one hospital and then tested with medical imagery from another hospital. 

In machine learning, within-distribution generalization is rather straightforward whereas out-of-distribution generalization is generally more difficult to accomplish. In image classification \cite{Recht2019-gi, Lu2019-cf} and robot manipulation \cite{Atkeson2018-du}, even small variations in the data (e.g., the way images are annotated) can lead to a significant drop in generalization performance. Out-of-distribution generalization to a new region in stimulus space is also challenging for humans and non-human primates, except when relying on explicit rules \cite{Casale2012-dx, Smith2015-hz}. Another approach to out-of-distribution generalization is transfer of conceptual knowledge of a learned relational structure to novel stimuli \cite{Kumaran2009-cn}. 

While the distinction between within-distribution and out-of-distribution test data is an important factor in evaluating generalization performance, it may not always be clear whether the test stimuli follow this dichotomy. This is particularly true in neuroscience and psychology studies where the distinction may depend on a variety of experiment-specific variables and goals.

Finally, although performance on training items themselves is never tested in machine learning, their inclusion is common and informative for human subjects. Interestingly, humans often rely on generalization processes even for the highly studied training items \cite{Shepard1961-ll}. This should be taken into consideration in neuroscience and psychology studies where comparisons are made between responses to trained and novel data. 

\subsection{Different terms for the same thing?}
While the term "generalization" likely means several distinct phenomena that are sometimes, perhaps mistakenly, referred to as equivalent, an equally important challenge is to determine which distinct terms refer to the same phenomenon. For example, “similarity-based” generalization (a term that we borrowed from Wasserman et al., 1992 \cite{Wasserman1992-ym}) may be equivalent to what others refer to as “primary” stimulus generalization \cite{Wasserman1992-ym, Hull1943-ts, Bialer1961-ca} or “feature-based” generalization \cite{Davis2017-fy, Wutz2018-wm}. Relational generalization (a term we borrowed from \cite{Davis2017-fy}) may be equivalent to what others refer to as “secondary stimulus generalization” \cite{Hull1943-ts, Bialer1961-ca}, “mediated stimulus generalization” \cite{Wasserman1992-ym}, “non-similarity-based generalization” \cite{Wasserman1992-ym}, and/or even “higher-order conditioning”. Relational generalization is itself a superordinate term for other frequently used terms, such as “transitive inference” \cite{Dusek1997-hf} and “associative inference” \cite{Preston2004-tq}. Proposed mechanisms may also have multiple names. For instance, the term “integrative encoding” \cite{Shohamy2008-qd} and “retrieval mediated learning” \cite{Iordanova2011-ex} are referring to the same underlying mechanism of relational generalization. Having multiple terms for the same concepts makes synthesis and cross-talk challenging. There is no easy way to harmonize terminology across labs and fields of study, because certain terms may be well established in a given subfield and an attempt to harmonize with another field of study may cause confusion. By establishing cross-lab interdisciplinary collaborations, such as general adversarial collaborations, common ground may be more readily identified.	

\subsubsection{Should we really call all these phenomena “generalization”?}
While each of the examples noted above is considered a form of generalization by several research groups, there are many that reserve the term “generalization” for what we refer to as “similarity-based generalization”. For example, the definition of generalization by Mazur (2016) \cite{Mazur2015-pg}, typically abided by researchers from the field of animal learning, specifies that generalization occurs when a response to a conditioned stimulus is extended to a similar stimulus (like in our example above about Pavlov’s dogs responding to tones of different frequencies even though learning included only a single frequency tone). This definition by Mazur (2016) extends beyond perceptual generalization to include generalization across conceptual similarity space, such as when consumers sometimes extend their favorable attitude about a product with a given brand name to other products with the same brand name (based on the work by Till \& Priluck, 2000 \cite{Till2000-vz}). Thus, generalization according to this definition includes multiple forms of similarity-based generalization, but not what we refer to as relational generalization (with rule-based generalization remaining ambiguous). Focusing on similarity-based generalization provides the benefit of clearer assumptions regarding the underlying neural mechanism. For example, we know that the neural representation of a 400 Hz sound would be more similar to a 350 Hz sound than a 200 Hz sound, which explains stronger generalization from 400 Hz to 350 Hz than 200 Hz. One can also use measures of behavioral generalization to make assumptions about the relative similarity of neural representations for different stimuli. Researchers who follow this stricter definition of generalization often refer to the phenomenon that we refer to as “relational generalization” by the broad term “transfer” (this also applies to rule-based generalization) or by more specific terms, such as ”higher-order conditioning” or “sensory preconditioning”. 

Given that “generalization” in the narrower sense of similarity-based generalization seems more precisely defined, shouldn't we all adopt the more conservative definition rather than lumping relational generalization into the same bag? It is not that simple. As discussed later in this paper, similarity-based generalization (generalization in the narrow sense) and relational generalization (such as transfer or inference) might actually share conceptual and neural mechanisms. For now, however, we acknowledge that what counts and does not count as generalization is not universally agreed upon.	

\subsection{Different experimental subjects}
To date, a variety of different experimental subjects have been used to investigate generalization, including rodents, pigeons, monkeys, and humans \cite{Barron2013-iy, Bowman2018-pd, Barron2020-bm, Honey1989-lf, Honey1990-xl, Maes2015-yl, Murphy2008-uw, Pan2008-jm, Pan2014-lx, Bowman2020-hv}. The type of experimental subject selected in each study is often determined by the methodology required to answer the specific question(s) of interest. Questions at the cellular level require more invasive techniques (e.g. single-cell recordings) and have therefore been largely restricted to animal models. These invasive techniques provide excellent spatial and temporal resolution but cannot currently be used to simultaneously record activity from the whole mammalian brain. Instead, recordings are typically made from one or two target brain regions of interest, whose selection is constrained by the general focus of a given lab and practical limitations such as ease of recording from a particular site. Whether signals observed in the target region of interest may be also represented elsewhere is typically unknown.  

In contrast, questions at a more macroscopic level are generally studied in humans, utilizing non-invasive techniques that can record activity from the whole brain (e.g. fMRI and EEG). Functional MRI, in particular, has been one of the most fruitful neuroimaging techniques in humans, providing novel insights into the network organization of the brain and increasing our understanding of functional neuroanatomy and neural representations. However, non-invasive whole-brain techniques typically rely on indirect measures of neural activity and may prioritise spatial (fMRI) or temporal (EEG) resolution, but cannot achieve both simultaneously. Finally, even “high-resolution” fMRI includes thousands of neurons within each voxel (spatial unit of measurement). 
				
The different focus in different species has produced some encouraging converging results, but also left several white spots on the map. While some brain regions, such as the hippocampus, have been investigated in the context of generalization in both humans and animals, other regions firmly established within one species are yet to be investigated across species. For example, the ventromedial prefrontal cortex (VMPFC) has been implicated in a number of human studies on generalization \cite{Zeithamova2012-ut, Kumaran2009-cn, Schlichting2016-co, Van_Kesteren2013-yh, Warren2014-gw} and even in some animal lesion studies \cite{DeVito2010-tf}. However, single cell recordings from this region are limited by lack of homology across species and by practical difficulties. By contrast, regions within the lateral prefrontal cortex (LPFC) have been frequent regions of interest in animal cognitive research (e.g., landmark studies by Earl Miller, Patricia Goldman-Rakic, and others). This includes work on some forms of generalization \cite{Wutz2018-wm, Freedman2001-zu}, even though the LPFC is not widely regarded as a brain region supporting generalization in human subjects. This discrepancy between species may, in part, reflect differences in tasks or differences in brain anatomy between species \cite{Neubert2014-fu, Neubert2015-ja}, together with differences in the methods being used. Finally, even human fMRI analyses sometimes focus on predetermined regions of interest. This reduces the number of investigated voxels and therefore the severity of the required corrections for multiple tests, but may miss other regions beyond those hypothesized a priori. 
					
Lastly, it is important to consider experimental tasks themselves, which need to be tailored to the experimental subjects. The key differences include (1) simpler tasks being typically used in experiments with animals compared to humans, (2) use of verbal instructions in humans, (3) prolonged learning across many repetitions and often many days or weeks in animals compared to a few repetitions or even single shot learning in humans, (4) the need to use reward or other behaviorally salient components to achieve a behavioral read-out of learning in animals, which is not always necessary in humans. Each of these task features -- difficulty, instruction, repetition, motivation -- has been documented to affect learning and behavioral and neural measures, complicating our ability to conduct cross-species comparison. Together, these factors have resulted in a somewhat biased literature, where (a) simpler generalization tasks are typically investigated in animals at the cellular level, and (b) more complex generalization tasks are typically investigated in humans at a coarser, whole brain level. Because of the inevitable differences in task and methods, it is challenging to discern to what degree any discrepancies between findings in different species are driven by differences in task and methods, and to what degree they reflect true underlying differences in the mechanisms involved in humans versus other species. Bridging this gap will be necessary for a more complete understanding of generalization.	

\subsection{Different task designs}
We have already discussed that generalization is a heterogeneous concept that encompasses several different phenomena. For example, both similarity-based generalization and relational generalization have been operationalized using many different tasks. The degree to which these different tasks measure the same underlying function is not always clear.		

\subsubsection{Episodic inference tasks}
Episodic inference tasks constitute one family of generalization tasks that clearly focus on relational generalization. Notably, these tasks are frequently considered to measure the same underlying relational generalization process \cite{Zeithamova2020-sh}. The most commonly used episodic inference tasks -- acquired equivalence \cite{Shohamy2008-qd, Honey1989-lf}, associative inference \cite{Zeithamova2012-ut, Barron2020-bm, Preston2004-tq}, and transitive inference \cite{Heckers2004-ri, Ryan2016-xk, Zalesak2009-yf} -- are depicted in Figure 1. A real-world example of an acquired equivalence might be that you learn that Barry and Susan both like to go hiking. After learning that Barry also likes rock-climbing, you assume that Susan probably does too. A real-world example of associative inference might be that after seeing Dave in front of a given house and then later seeing Jan in front of the same house, you make the assumption that Dave and Jan are relatives or partners who live there together. One exhibits transitive inference when inferring that pineapples are more expensive than oranges, after learning that pineapples are more expensive than kiwifruit and that kiwifruit are more expensive than oranges. Experimental tasks designed to induce all of these types of episodic inference have been examined in multiple species, with some exciting convergence, particularly regarding the role of the hippocampus and mPFC in both humans and rodents \cite{Preston2013-kd} However, the tasks typically have to be adapted before they can be used in different species, as illustrated in Figure 1 with the human (Fig. 1b) and animal (Fig. 1c) versions of the associative inference task (train A-B, B-C, test A-C). Human subjects typically learn a large number of overlapping associations by simply observing pairs of stimuli on a computer screen, each pair briefly presented 1-3 times. Which response is expected for the inference test trials (A-C) is usually explained to the participants. This, as well as the overlapping pair structure, may even be revealed before learning begins \cite{Zeithamova2010-ux}. In contrast, a typical rodent version \cite{Bunsey1996-gm} contains just two sets of overlapping associations (A-B, B-C and X-Y, Y-Z) that are trained over many repetitions spread out over several days. An animal learns a choice for each cue (e.g., to choose odor B over odor Y when presented with the cue odor A) through reward feedback, with no possibility of explicit verbal instructions. Synthesis of human and animal research needs to take those differences into consideration because the number of repetitions, opportunity for sleep/rest, explicit instruction, and the presence of reward have been all shown to affect learning. Some studies have attempted to address this by using reward in both animal and human generalization tasks and avoiding explicit inference instruction in humans \cite{Barron2020-bm}, helping to equate at least some task features across species and facilitate comparison.		

\begin{figure}[H]
\centering
\includegraphics[width=9cm]{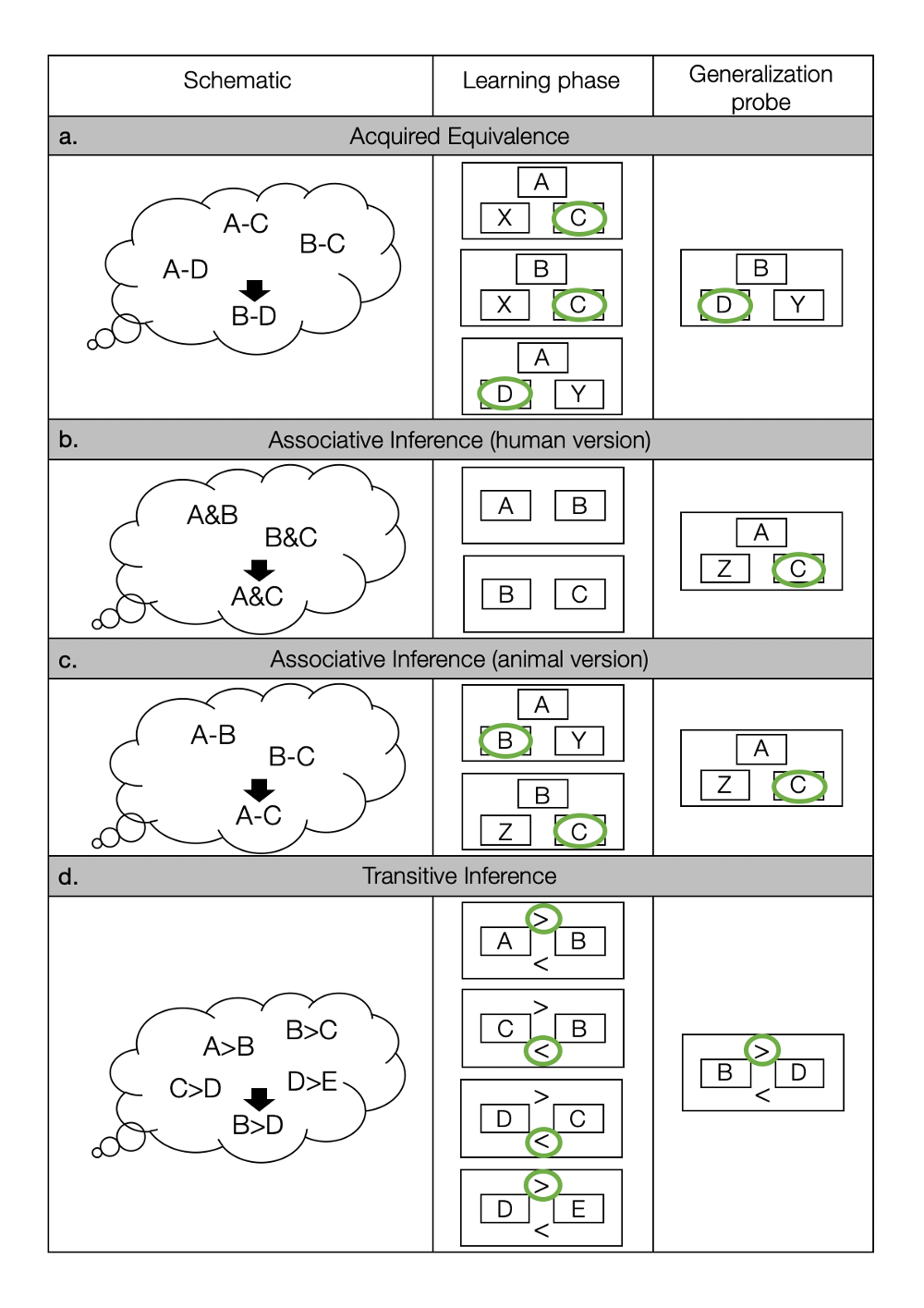}
\caption{Figure 1. Episodic inference tasks. Letters A, B, C, D, E stand for specific stimuli (e.g., an image on the screen for a human, an odor for a rodent). Green circles represent the correct answers. a. In Acquired Equivalence tasks, participants learn that two probe stimuli are paired with the same choice (A-C and B-C), and also that one of the probe stimuli is paired with another choice (A-D). They are then tested to see if they choose D when presented with B (B-D). b. In human versions of associative inference tasks, participants encode overlapping paired associates (A\&B and B\&C) via observational learning. They are then tested on the relationship between A and C. Most commonly, subjects are presented with an A probe and asked to select between two stimuli, with the choices being the correct C item and another familiar item. c. In the rodent versions of associative inference tasks, subjects learn through reinforcement learning over many trials to choose B over Y when presented with A, and choose C over Z when presented with B (A-B, B-C). They are then tested to see if they will choose C over Z when presented with A, although the A-C choice was never trained. d. On each trial of transitive inference tasks, participants are shown two images on the screen and asked to choose one. Over many trials through corrective feedback, they learn to choose A over B, B over C, C over D and D over E (A>B, B>C, C>D, and D>E). They are tested to see if they will generalize to assume, for example, B>D.}
\end{figure}

\subsubsection{Traditional categorization tasks}
A second group of generalization tasks are categorization tasks that require a subject to learn a common label or response to denote that a set of stimuli belong to the same category. As the key generalization aspect, categorization tasks typically require generalization of the categorization response to new stimuli, for which the correct response was not explicitly trained. Most categorization tasks elicit similarity-based generalization, although we will discuss some exceptions later on. It should be noted that even categorization tasks that are similarity-based may rely on distinct mechanisms to one another. Categorization research has a long history and a number of stimulus types have been popular over the decades (Figure 2). Examples include dot-pattern stimuli \cite{Posner1968-hn}, which have adapted for monkeys \cite{Wutz2018-wm}; stimuli varying along two continuous-valued dimensions such as lines or Gabor patches \cite{Zeithamova2006-en}; cartoon-like drawings with several controlled binary-valued dimension \cite{Shepard1961-ll, Davis2012-ke, Bozoki2006-ej}; or even paintings by different artists where the experimenter does not have explicit control over the specific features relevant for categorization \cite{Watanabe1995-jo}. Categorization stimuli may also include stimulus-specific features and/or unique identifiers (names) to allow measuring both generalization across stimuli and memory for specific stimuli in the same task (e.g., cartoon satellites \cite{Schapiro2017-nb}; blended human faces \cite{Ashby2020-ok}).

\subsubsection{Challenges in comparing generalization in episodic inference tasks and categorization tasks}
Generalization in traditional categorization tasks is usually similarity-based. On the other hand, generalization in most episodic inference tasks is relational. Therefore, it would be tempting to assume that traditional categorization tasks measure one type of generalization relying on one mechanism while episodic inference tasks measure another type of generalization relying on another mechanism. However, there is substantial evidence that there exists (a) several distinct category generalization mechanisms, and (b) a substantial overlap between mechanisms of relational generalization and categorization.
					
First, a body of research indicates that in traditional categorization tasks the mechanism can vary depending on the category structure and upon the task itself (see \cite{Gregory_Ashby_undated-rh, Ashby2005-es} for a review). For example, the same category structure can be easy or hard to learn and lead to less or more storage depending on the induction task (e.g., inference versus classification \cite{Love2004-jl}. Some of the dissociation proposed and empirically demonstrated include generalization based on the overall similarity versus a unidimensional verbalizable rule \cite{Maddox2004-kc}; learning characteristic features of a single category versus learning two contrasting categories \cite{Zeithamova2008-rk}; and incidental perceptual grouping versus intentional instructed category learning \cite{Aizenstein2000-nn}. For example, some studies have A/not-A structure (Fig. 2a) where subjects are first exposed to examples of category A and then tested to see whether they can generalize information such that they are able to differentiate between new examples of category A and stimuli that are not members of category A. Other studies have A/B structure where subjects first learn to differentiate members of two or more categories and are then tested to see whether they can then generalize the learned category labels to new stimuli. Existing work indicates that A/not-A learning may be preserved in amnesia and rely on the striatum, while A/B learning may instead rely on the hippocampus \cite{Knowlton1993-xf, Zaki2003-oq, Bozoki2006-ej, Zeithamova2008-rk}.
					
Interestingly, although research on episodic inference and categorization research have proceeded somewhat in parallel, recent work indicates that some forms of categorization may actually rely on the same mechanisms implicated in episodic inference, including hippocampal-based memory integration \cite{Zeithamova2020-sh, Bowman2018-pd}. Furthermore, the same task seems to sometimes rely on memory integration and at other times rely on the encoding and retrieving of specific category exemplars \cite{Bowman2020-bp}, depending on within-task conditions. Thus, the best way to classify different generalization tasks and the degree to which superficially similar tasks actually measure the same underlying process remains an open question.

\begin{figure}[H]
\centering
\includegraphics[width=9cm]{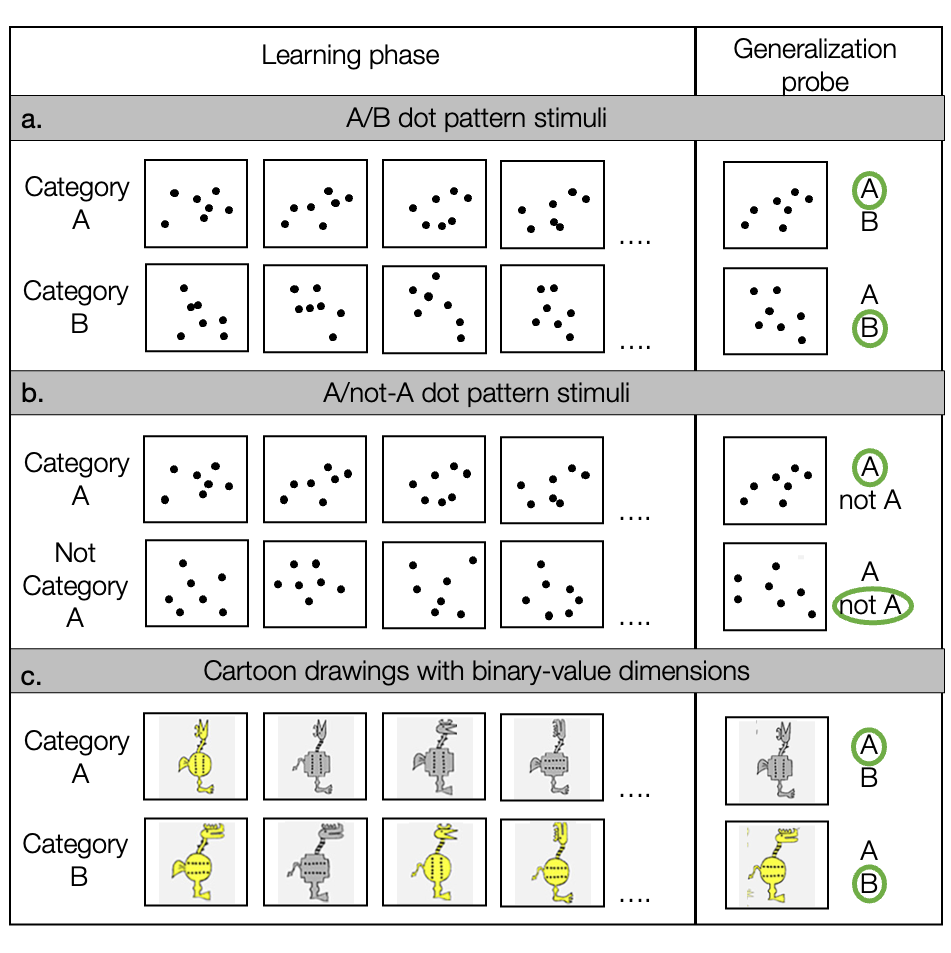}
\caption{\textbf{Traditional categorization tasks}. Green circles represent the correct answers. a. In the A/B dot pattern stimuli task, via feedback, participants learn to categorize two different groups of dot pattern stimuli (A and B). Each of these categories contain stimuli that share a central tendency, created as distortions from a prototype pattern by moving individual dots with a controlled amount of jitter. Participants are subsequently tested to see if they can accurately categorize: the stimuli from the learning phase; prototypical stimuli from each category; and new stimuli from each category \cite{Posner1968-hn}. b. The A/not-A pattern stimuli task only includes examples from a single category during training. At test, new categorical examples are presented as well as new non-categorical examples that are typically random patterns that do not share a central tendency with stimuli from the category or with each other \cite{Knowlton1993-xf}. c. An example of a categorization task that uses stimuli with well- defined binary-value dimensions. Here, cartoon animals vary based on binary characteristics such as color (e.g., yellow/grey), body shape (squared/circular), and/or head orientation (forward/up). Category membership may rely on a single dimension, two dimensions, or more dimensions \cite{Shepard1961-ll}. Some versions of this type of categorization task are analogous to the dot pattern stimuli task. In this case, one stimulus may serve as the category A prototype and category A exemplars are constructed as its “distortions”, where some of the category-typical features are exchanged for atypical features (stimuli in this figure were adapted from Bozoki et al., 2006 \cite{Bozoki2006-ej}).}
\end{figure}

\subsubsection{Non-traditional categorization tasks and the “functional category reward prediction task”}
Categories do not have to be based on perceptual similarity. For example, Barsalou (1983) noted that categories can be determined by conceptual similarity, such as the category “fruit” which contains the perceptually distinct but conceptually similar “apple” and “banana” \cite{Barsalou1983-ca}. Category membership can also be determined in a more ad-hoc fashion. For example, although an apple, a tent, and a lighter are not perceptually or even conceptually similar at first, they may begin to gain conceptual similarity after being classified into the same items- to-bring-camping category \cite{Barsalou1983-ca}. Markman and Stilwell (2001) highlighted the existence of role-governed categories, such as the category “nurse”, which is defined by a specific job/ specific role in society \cite{Markman2001-tn}. Jones and Love (2007) noted categories defined by relationships with other stimuli \cite{Jones2007-wx}, such as the category “predator” being defined by the role an animal plays during hunting. Category membership may also be determined by function \cite{Pan2008-jm, Pan2014-lx}, such as when a \$10 bill, \$10 in coins, a check for \$10, and \$10 on your credit card will all function equally well in terms of purchasing power (we shall refer to such categories as “functional categories” for short).

In the context of generalization, we will specifically focus on the last example, which was adapted for the single-cell recording studies in monkeys by Pan, Sakagami and colleagues using their “functional category reward prediction task” (Figure 3, \cite{Pan2008-jm, Pan2014-lx}). Despite including “category” in the name, this task differs substantially from the traditional category generalization tasks described in section 2.3.2. In its simplest form, monkeys first learn that a set of perceptually and conceptually unrelated stimuli form a category. Despite not sharing any initial similarities, monkeys learn that members of a category share “functionality” in that they all always function to predict the same amount of reward (this amount changes from session to session but is always consistent for members of a given functional category). Because the categories are arbitrary, one cannot test generalization of category labels to new stimuli. For instance, it is not possible to figure out which functional category a book belongs to after learning that a triangle, an apple and a shoe are all members of functional category 1, while a lighthouse, a chair and a daffodil are all members of functional category 2. Instead, one can only test for transfer (or what we call relational generalization), such as generalizing a piece of information learned about one category member to another member. 

In Pan et al., (2008; 2014), after the initial category training, one member of category 1 was associated with high reward and one member of category 2 was associated with low reward (or vice versa). Neurons in the striatum and lateral prefrontal cortex (LPFC) were recorded while the monkey was presented with other members of each category. A set of LPFC neurons were found to respond selectively to stimuli associated with their preferred reward level. Another set of LPFC neurons was found to respond selectively to stimuli from their preferred stimulus category. In evidence for the transfer of reward information between members of a given category, yet another set of LPFC neurons were found to combine reward and category information so that they respond more strongly to stimuli associated with their preferred reward level only if they were from their preferred stimulus category (e.g. only to stimuli from category 1 and only when these were associated with high reward). This was found even for category members that themselves had not been directly experienced with reward before (as long as another member from the same category had been), providing evidence for the generalization of reward expectation from one category member to another. 

Notably, the striatum did not show reward predictive information for new category members the first time that they were presented, but did so on subsequent test trials with the same stimuli. This indicates that once the generalization response has been computed during the first encounter, it can simply be retrieved on subsequent repetitions and may be represented by regions not involved in its initial computation. Thus, caution is necessary when interpreting a lack of neural activity unique to generalization trials (as opposed to direct retrieval trials) in studies that repeat generalization probes \cite{Shohamy2008-qd}.

Besides Pan et al., several other studies have evidenced the role of LPFC neurons in coding functional categories. Using perceptual categories, EK Miller and colleagues demonstrated that LPFC neurons represent flexible changes of a category border, dependent on the changes in behavioral meaning \cite{Wutz2018-wm, Freedman2001-zu}. Hoshi and Tanji (2000) showed that when monkeys are trained to touch a left panel with their left hand and a right panel with their right hand in one task condition and to touch the left panel with their right hand and the right panel with their left hand in another task condition, M1 and Premotor neurons discriminate left/right hand use while instead LPFC neurons discriminate left/right target (panel) \cite{Hoshi2000-xl}. This suggests that LPFC neurons code more abstract signals based on the integration of sensory inputs and feedback from output (motor) signals, which may thus generate functional categories.

\begin{figure}[H]
\centering
\includegraphics[width=9cm]{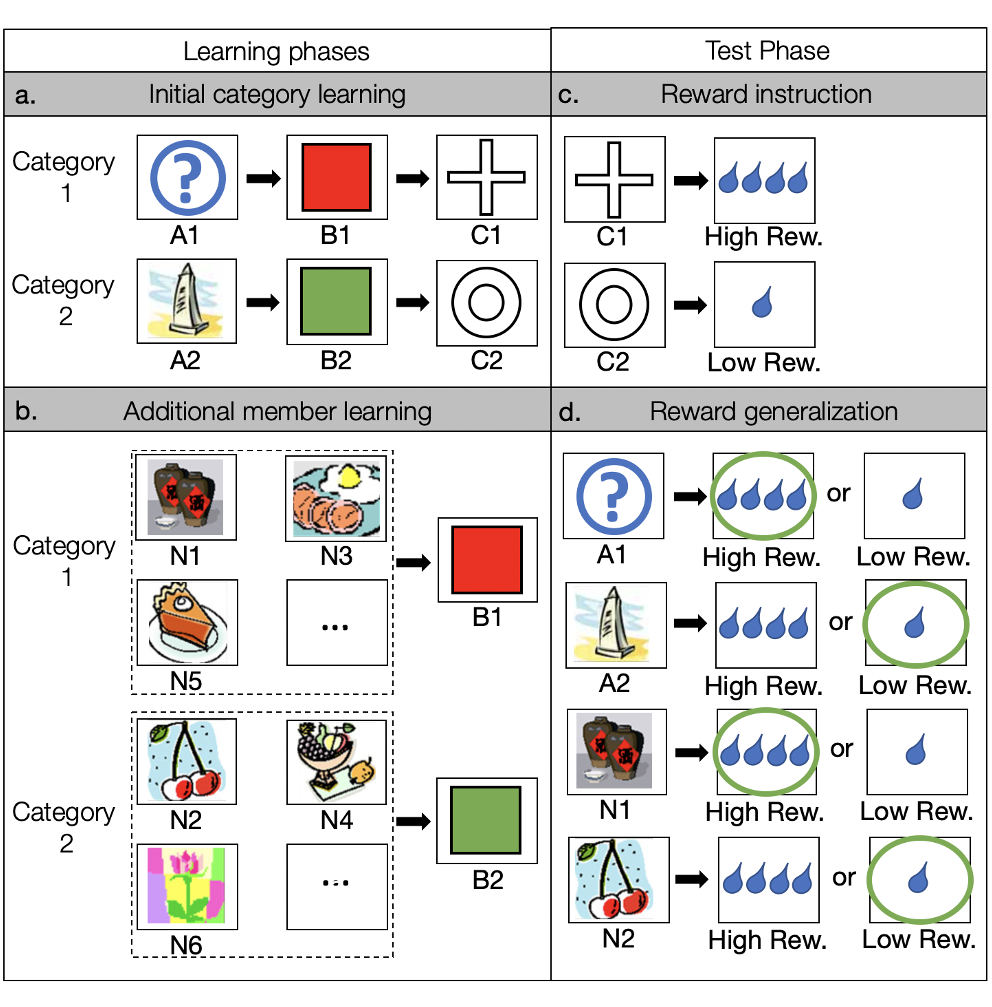}
\caption{Figure 3. The functional category reward prediction task of Pan et al. (2008; 2014). Green circles represent the correct answers. a. During the initial category learning phase, via trial and error, subjects learn to associate groups of stimuli with one another to form two functional categories. If, after seeing one stimulus from a given category (e.g. A1), the subject next selects two other stimuli from that category (e.g. B1 and C1) then they receive a reward. If they do not select stimuli from the same category (e.g. choose B2 and/or C2 after seeing A1) then they are not rewarded. Although not depicted here, subjects learn the stimuli from each group in several different temporal sequences (e.g. A1-B1-C1, B1-C1-A1, and C1-A1-B1). This is to ensure that symmetry is established between category members. b. After initial categories have been well learned, in a separate learning phase, subjects learn to associate multiple novel stimuli with a stimulus from one of the categories (e.g. B1) and multiple other novel stimuli with a stimulus from the other category (e.g. B2). c. At the beginning of each session of the test phase, participants complete several reward instruction trials. In these trials, participants are presented with a stimulus from one category (e.g. C1) along with high reward, and a stimulus from another category (e.g. C2) with low reward. d. After the reward instruction trials in each session of the test phase, participants are tested to see if they will transfer the reward information learned during the reward instruction trials to other members of the same functional category; This is tested for both category members learned during the “initial learning” phase (e.g. to see if they combine A1-B1, B1-C1, and C1-large reward to begin predicting large reward to A1) and for novel category members learned later in the “additional member learning” phase (e.g. to see if they combine N1-B1, B1-C1, and C1-large reward to begin predicting large reward to N1) . Rew. = reward.}
\end{figure}

\subsubsection{Challenges in comparing generalization in functional categories, similarity-based categories, and other generalization studies}
Some of the challenges in understanding how the research on functional categories from Pan and colleagues \cite{Pan2008-jm, Pan2014-lx} fits with more classical generalization research stem from (a) the lack of a human task equivalent, (b) having to limit single cell recordings to specific regions of the brain, and (c) the unclear position of where tasks that use functional categories fit on the spectrum of generalization tasks. For example, Pan and colleagues (2008, 2014) found neurons in two regions in the brain-- the striatum and the LPFC-- that store information at two different levels (at the individual stimulus level in the striatum and at the category level in the LPFC), but these were the only two regions that were measured in these studies. How comparable, therefore, are these results to those of other generalization studies?
					
Because functional categorization tasks elicit relational generalization rather than similarity-based generalization, at a conceptual level functional categorization tasks can be considered more similar to episodic inference tasks than traditional categorization tasks. In fact, functional categories like those used by Pan and colleagues could be considered to be an extended case of associative inference (Figs. 1b,c \& 3a,b). Nonetheless, studies on associative inference and functional categorization rarely reference each other and focus on different regions of interest. One reason may be that researchers who study associative inference in humans would view it as a form of generalization, whereas functional categorization does not fit into the definition of “generalization” within the field of animal learning.		
					
The closest study in humans to the functional category task used by Pan and colleagues may be a study on the associative inference of reward by Wimmer \& Shohamy (2012) \cite{Wimmer2012-te}. This, however, highlighted the role of the hippocampus rather than LPFC in the generalization of reward information. It is not possible to discern to what degree these discrepancies are driven by different a priori regions of interest (hippocampus versus LPFC) and to what degree they stem from other task differences (e.g. overlearning of the categories in monkeys) and analysis differences (e.g. averaging responses across multiple presentations of the same stimulus \cite{Wimmer2012-te} versus specifically examining the first response to each stimulus \cite{Pan2008-jm, Pan2014-lx}).
					
While papers on functional categories often highlight the non-perceptual nature of their categories, what role this plays in determining the underlying category generalization mechanism is not clear. For example, many studies that used perceptual rather than functional categories in humans or animal models have also highlighted the LPFC in representing abstract category information and in supporting generalization \cite{Davis2017-fy, Wutz2018-wm, Freedman2001-zu, Mack2013-mj, Mok2020-pv}, so this neural region does not appear unique for non-perceptual categories. Because members of a given category may become conceptually similar to one another during learning \cite{Barsalou1983-ca, Tomlinson2006-fr}, it is plausible that generalization in functional categories and perceptual categories relies on the same, similarity-based principles. Furthermore, other studies using direct recordings in primates have instead pointed to the posterior parietal cortices as loci of abstract representation of category information \cite{Zhou2019-eh} highlighting the challenge of comparing studies with distinct a priori ROIs. Overall, how functional categorization tasks and their results fit into the larger picture of the generalization literature needs to be better determined. Within our GAC, we all see functional categories as relevant to our understanding of generalization, but are not unanimous in where we perceive this task to sit with respect to other generalization tasks.

\section{More than one mechanism may be at play}
On the one hand, an array of generalization behaviors have been demonstrated. On the other hand, several proposed mechanisms of generalization have been proposed. It may thus be tempting to assume that different generalization mechanisms perhaps underlie different types of generalization behavior. However, the situation appears to be more complicated than a set of one-to-one mappings between distinct behaviors and mechanisms. Recent work suggests that more than one of the proposed generalization mechanisms may give rise to the same generalization behavior and that more than one type of generalization behavior may be explained by one of the proposed generalization mechanisms. While a more thorough discussion can be found elsewhere \cite{Zeithamova2020-sh}, here we will note some of the current findings that indicate multiple mechanisms at play, often within the same task. We will also note the potentially surprising parallels across different tasks. Testing these ideas may be one of the first targets for empirical research and computational models of generalization in the near future.

\subsection{Both memory integration and on-the-fly generalization may contribute to both episodic inference and category generalization}
The current Generative Adversarial Collaboration started with the goal to resolve differences between a memory integration view of generalization and the view that generalization is accomplished on-the-fly based on separate specific memories. In the process, we have discovered that even though particular papers of ours may argue one way or another, nearly all of us are actually in agreement that both mechanisms likely support generalization, under different conditions, in different tasks, in different subjects, and that they may even operate in parallel.	

In episodic inference studies, compelling evidence was found both for integrative encoding \cite{Shohamy2008-qd} and for retrieval-based on-the-fly inference based on separate memories \cite{Barron2013-iy}. Furthermore, some studies have shown evidence for both integrative encoding and on-the-fly generalization in the same task \cite{Zeithamova2010-ux, Schlichting2016-co, Zeithamova2017-ut}, indicating that these mechanisms may not be mutually exclusive. Interestingly, we would like to argue that both integrated and separate representations seem to play a role in the generalization of categorical knowledge as well. The assumptions of integrated versus separated representations are embedded in the prototype and exemplar models of categorization, respectively. Whether generalization is better explained by assuming it is based on an exemplar representation of specific category members \cite{Nosofsky2000-fl}, or a prototype representation that constitutes a summary representation abstracted across category members \cite{Smith1998-fe}, has been debated for decades. 

\subsubsection{Prototype and exemplar models as tools for indexing integrated versus separate representations}
In prototype models, category learning results in the formation of a generalized category representation. This is the ideal member of a category or an integrated average of existing members of a category, called a “prototype”, which guides generalization judgements for new stimuli \cite{Bowman2018-pd, Posner1968-hn, Smith1998-fe}.  In exemplar models, categories are represented by all previously encountered members of a category, called “exemplars”, that jointly guide generalization judgements for new stimuli \cite{Nosofsky2000-fl, Mack2013-mj, Kruschke1992-yj, Medin1978-ex}. Both of these types of models assume that one compares the current stimulus to representations (prototype or exemplar) of all relevant categories and chooses the category that is perceived as the most similar, i.e., closest to the current stimulus in perceptual space. Because the relationship between physical similarity and perceived similarity is assumed exponential rather than linear \cite{Shepard1957-xl}, perceived similarity of a new stimulus to the category prototype is not mathematically equal to its perceived similarity to the exemplars, resulting in dissociable predictions from the two models under most circumstances. 

Exemplar models, which assume that categories are represented by the specific instances that a subject encountered during training, account well for performance in many laboratory categorization tasks. However, what is an “exemplar” in the real world context is ambiguous \cite{Love2004-jl}. For example, one may consider “the neighbor’s dog” to be a single exemplar of the category “dogs”, or one may consider each encounter with the neighbor’s dog to be stored as an exemplar. When passing by a new dog, multiple unique images of that dog with slightly different views are projected onto our retinas. Do we store a new exemplar for each of these or just one new exemplar as an integrated average across time and viewpoints? In other words, what is an exemplar at one level of description may be itself a “prototype-like” abstraction from many exemplars at a higher level. The categorization model SUSTAIN circumvents this issue by not differentiating between prototypes and exemplars and instead assumes that similar experiences can be coded together as “clusters” at varying levels of abstraction \cite{Love2004-jl}. 

Despite their shortcomings, exemplar and prototype models have proven fruitful as indices of the representations that generalization judgments are based upon, respectively: relatively specific representations versus relatively generalized (integrated) representations. For example, a recent study found neural signals consistent with the prototype model in the hippocampus and ventromedial prefrontal cortex, the same regions shown to support integrated memories in episodic inference paradigms \cite{Bowman2018-pd}. A follow-up study showed that both prototype and exemplar representations of categories can emerge in parallel in the brain during learning \cite{Bowman2020-hv}, providing further evidence that generalization mechanisms based on specific and integrated representations might not be mutually exclusive. Finally, several studies replicated the same hippocampal-VMPFC interactions in prototype-based categorization \cite{Bowman2018-pd, Frank2019-wx,  Bowman2020-hv} and in memory integration of overlapping relational memories in episodic memory tasks \cite{Zeithamova2012-ut, Schlichting2016-co}. This overlap in principles and neural mechanisms indicates that superficially very different tasks may rely on the same sets of mechanisms. 

\subsubsection{Neural dissociations that may map onto separate versus integrated memory codes}
Supporting the idea that multiple representations can be stored in parallel in the brain is growing evidence for a topological divide in hippocampal function: coarser, integrated information is represented in the anterior hippocampus and more fine-grained, specific information is represented in the posterior hippocampus (see \cite{Poppenk2013-uy} for a review). Human fMRI further indicates the concurrent presence of both types of representations \cite{Schlichting2015-rj, Bowman2020-hv, Brunec2018-uk, Collin2015-ky}. The human anterior/posterior distinction parallels the ventral/dorsal gradient in rodents, with coarser, larger receptive fields in the ventral hippocampus and finer, smaller receptive fields in the dorsal hippocampus \cite{Kjelstrup2008-he}. A similar topological divide is found along the medial-lateral axis of the human PFC, with the mPFC representing more integrated and the LPFC representing more specific information \cite{Bowman2018-pd, Bowman2020-hv, Mack2013-mj}. However, even this may be an oversimplification, as can be seen from studies showing both integrated and specific representations within the mPFC \cite{Schlichting2015-rj} and within the LPFC \cite{Wutz2018-wm}. 

\subsubsection{Challenges to inferring underlying representations}
As noted earlier, different pathways may lead to the same behavioral outcome. Many of the recent advances which allow better arbitration between these mechanisms result from advanced analysis and modeling approaches, such as multivariate pattern analyses or utilizing quantitative predictions of mathematically formalized prototype and exemplar models in fMRI data analysis. However, estimating the underlying representations from patterns of behavioral and neural data always involves an inference and relies on assumptions that some may call into question. For example, could a poor memory for specific exemplars manifest as a prototype representation? When someone’s behavior indicates that they are focusing on a single dimension of a multidimensional stimulus, should we conclude they are using an explicit rule \cite{Maddox2004-kc}? Or should we treat it as an exemplar representation with strong selective attention \cite{Nosofsky1986-pb}? And can we rule-out a prototype representation if a prototype model provides an equally good data fit? To be confident in our interpretations, assessing the same questions using complementary measures and converging evidence should be beneficial. 

\subsection{Conditions that bias integrated versus separate encoding of information}
In the prior section, we discussed evidence that events may be represented at multiple levels of specificity simultaneously. However, some studies have observed generalization behavior to be dominated by primarily one mechanism, either integration or on-the-fly retrieval of separate memories. What may determine how we encode related experiences? 

\subsubsection{Similar experiences are likely integrated, dissimilar separated}
Increased similarity between related experiences seems to increase the probability of integration. In categorization tasks, a key factor in determining which type of representation one will form for the category is the degree of within-category similarity or coherence of training instances \cite{Wutz2018-wm, Bowman2020-bp}. When category members have high perceptual coherence, then it is easier to extract the central tendency of the category and category label generalization tends to be prototype-based. When perceptual category members have low perceptual coherence then category label generalization tends to be exemplar-based. The same categorization task may thus rely on distinct mechanisms, with dissociable behavioral and neural signatures \cite{Bowman2020-hv}. Similarly, in associative inference tasks, similar context promotes integration of overlapping associations (AB \& BC represented as ABC) while distinct contexts lead to encoding of overlapping associations as separate (AB, BC) \cite{Cohn-Sheehy2021-cm}.

\subsubsection{Events encountered close in time tend to be integrated in memory}
Another key factor shown to mediate the bias towards an integrated representation versus separate representation, identified in episodic inference tasks, is the temporal proximity in which events are experienced \cite{Zeithamova2017-ut, Cai2016-nt, Rashid2016-qa}. At the cellular level, the ensemble of neurons activated during the formation of a new memory do not immediately return to baseline levels of activity afterwards, so the same ensemble is more easily recruited for the formation of a subsequent memory (“memory allocation hypothesis” \cite{Silva2009-eq}). Representing events encountered close in time by overlapping neuronal populations leads to their integration and generalization of behavior from one event to the other \cite{Cai2016-nt, Rashid2016-qa}. Human fMRI studies have shown evidence consistent with this proposal, such as increased representational similarity for events encoded close in time \cite{Ezzyat2014-vc, Hsieh2014-go} and increased evidence for integration in an associative inference task when overlapping events occurred on the same day, as opposed to on two consecutive days \cite{Zeithamova2017-ut}. It may be the case, however, that while overall close temporal proximity of learning increases the chances that memories will be integrated, the first memory needs to be firmly established before the second memory can be integrated with the first. Consistent with this, Schlichting et al., (2015) found that overlapping events (e.g. A-B and B-C) were more likely to be integrated (to form an A-B-C memory) if these were learned in different blocks rather than in one interleaved block of learning \cite{Schlichting2015-rj}.

\subsubsection{Representations may change over time}
The studies discussed above indicate that representations of related events are sometimes encoded separately and sometimes encoded in an integrated manner. However, we also discussed in sections 1.3. and 1.4., some evidence that initially separate memories may ultimately become integrated through offline processing during sleep or rest or in response to task demands. For example, recalling multiple separate episodic memories may lead to their integration at the time when a generalization response is made, changing their representation from separate to integrated \cite{Carpenter2017-bi, Koster2018-sb, Sanchez2020-jw}. Thus, separation and integration pathways to generalization may ultimately converge to the same endpoint.

It is intuitive to assume that people must first store memories for specific events before they can connect the memories and generalize. However, the temporal sequence does not always evolve from separate memories to generalized memories. For instance, research about memory development has demonstrated that children build generalizable knowledge and conceptual understanding well before their memory for specific events matures \cite{Keresztes2018-pl}. One of our own studies, which allowed us to look for changes in representations underlying categorization decisions in a single experimental session, found no evidence of exemplars preceding prototypes during training \cite{Bowman2020-hv}. Moreover, a study by Johansen \& Palmeri (2002) found the opposite: reliance on generalized category information early in learning with exemplar retrieval only observed later in learning \cite{Johansen2002-hu}. Together, these results show that representations underlying generalization may change over the course of learning, time, or development, but that the direction of this shift is not straightforward and may depend on additional parameters.

\subsubsection{Conditions affecting other forms of generalization}
Going beyond the separation versus integration dichotomy, many other conditions and task factors have been shown to determine the type of representations and neural mechanisms that support generalization behavior. We have already mentioned the number of learning trials, verbalizability of a decision rule, feedback versus observational learning, A/B versus A/notA category structure, or the presence of reward. Additional highly studied factors include cognitive load \cite{Verguts2009-vw, Shanks1998-xs, Bright2014-hs, Wills2011-pb} or sleep \cite{Ellenbogen2007-se, Schapiro2017-tw, Pace-Schott2009-qi, Lau2010-eu, Schapiro2018-ig, Maddox2009-fj, Lerner2019-fj}. The degree to which the behavioral and neural differences found under these different conditions reflect entirely distinct generalization mechanisms versus differential engagement of a single common mechanism (e.g. better or worse online integration of information for generalization) remains to be clarified.

\section{The way forward in tackling the identified issues}
In this paper we have discussed how different generalization-related terms and different ways of operationalizing generalization are used within different subfields and by different groups of researchers. Cross-lab interdisciplinary collaborations, such as general adversarial collaborations, may provide a medium to unify these terms and inspire new approaches. Furthermore, a meta-analysis comparing results from existing studies that have investigated similar aspects of generalization (i.e. those suspected to represent the same underlying concepts and/or processes) could be of great benefit. Here, we outline three existing questions that we identified in the course of this Generative Adversarial Collaboration and that would facilitate better synthesis of findings across labs and species. After introducing each question, we provide a specific research study proposal to help answer the question.

\subsection{Question 1: Are we all measuring the same generalization phenomenon?}
A range of tasks have been used in research on generalization, but whether they capture the same underlying mechanism has not been evaluated. Different episodic inference tasks (Figure 1) are typically considered to address the same mechanism of memory integration. However, as we noted earlier, they all may involve various degrees of on-the-fly retrieval-based inference across separate memory representations. The episodic inference tasks also differ along several important dimensions -- feedback versus observational learning, number of repetitions, explicit instruction for inference versus uninstructed probes -- that have all been shown to affect the type of memory representations involved. It is thus important to determine the degree to which all these episodic inference tasks capture the same underlying generalization ability, dissociable from the ability to remember specific information.
					
Category learning has been studied largely in a separate line of research to episodic inference tasks. Because of the evidence that categorization tasks have provided for the multiple memory systems view \cite{Poldrack2008-de}, categorization has been often considered to rely on non-declarative forms of memory, such as the striatum rather than the hippocampus. Furthermore, any hippocampal contributions to category learning have been automatically assumed to signify memory for specific exemplars. Only recent work has started to demonstrate that prototype-based categorization may rely on the same hippocampal-based memory integration mechanism as proposed for the episodic inference tasks. Specifically, prototype formation and prototype-based generalization seems to involve hippocampal- VMPFC interactions that may be linking features of individual exemplars into integrated, prototype category representations \cite{Bowman2018-pd}. It is also possible that exemplar-based generalization relies on the same mechanisms as on-the-fly retrieval-based inference \cite{Zeithamova2020-sh}, which may or may not be fully accounted for by memory for specific episodes.
					
Individual differences provide a unique means to uncover the degree to which different tasks rely on shared or unique underlying cognitive processes. If the tasks measure a single generalization ability, then participant’s performance in one putative generalization task should track their performance in another putative generalization task (more so than other cognitive tasks). But do a participant’s behavioral and neural measures (response times, accuracy, neural response, or other) in one generalization task predict their results in other generalization tasks? Do subjects display reproducible neural activity across these tasks that would be predictive of generalization performance? We believe that examination of these questions is important to address the issue of whether or not we are all measuring the same generalization mechanism.
					
As a first step, we propose to employ a large N individual differences study that would test participants in a range of putative memory generalization and memory specificity tasks, in addition to control tasks that do not involve memory but require general motivation and compliance. Importantly, memory generalization tasks would include both episodic inference tasks and categorization tasks to test for their shared and unique variance. Using factor analyses, we would be able to test the hypothesis that individual differences in generalization across a range of tasks can be captured by one shared factor, dissociable from the ability to remember specific details. We will compare it to a model with two generalization factors, such as what could happen if some tasks primarily rely on memory integration while other tasks primarily rely on retrieval of separated specific memories. A subsequent latent-variable fMRI study would aim to link individual differences in memory specificity and generalization factors to resting- state and task-based fMRI metrics (connectivity, stimulus-evoked activation patterns) from the hippocampus and other regions previously shown to track performance in individual tasks. This work would determine to what degree different tasks capture the same underlying generalization ability or abilities, dissociable from the ability to remember specific information. Results will inform theoretical synthesis of existing work and future research directions, as well as provide a critical step toward the formulation of a comprehensive theory of generalization. Neuroimaging results will provide new means to dissociate hippocampal contributions to generalization through its role in memory for individual events and the postulated distinct role in forming representations that link information across events. The results will help resolve between theories of spatial segregation of distinct hippocampal codes across subfields and anterior-posterior axis.	

\subsection{Question 2: How does functional categorization arise in humans?}
Categories do not need to be based on the perceptual similarity of their members. “Functional categories” are one such example (see section 2.4.4.). Work in primates has demonstrated generalization of information between members of a functional category (Pan et al., 2008), but the underlying mechanisms remain to be examined in humans. 
					
To what degree do mechanisms of functional categorization align with those of episodic inferences (such as associative inference or higher-order conditioning) and those of perceptual categorization? On the one hand, distinct neural processes have been implicated for episodic inference (hippocampus, VMPFC) and functional categorization tasks (LPFC), despite these both involving relational generalization. On the other hand, common findings are sometimes reported for perceptual and functional categorization (e.g. the LPFC is often implicated in both in monkeys), despite the fact that perceptual categorization involves similarity-based rather than relational generalization. This may indicate a common mechanism behind categorization, regardless of whether it is perceptual or functional, as previously proposed for similarity-based versus relational categories \cite{Tomlinson2006-fr}.  Overall, because research on functional categorization, perceptual categorization, and episodic inference have largely evolved independently, it is difficult to determine to what degree the aforementioned differences and commonalities result from the differences in a priori regions of interest, the species used as subjects, and/or the tasks employed. We identified this particular problem as a starting point for our own future across-lab collaboration because the members of our generative adversarial collaboration have expertise spanning these different lines of research.	
								
Given that i) functional categorization appears related to both episodic inference and perceptual categorization, and ii) a variety of neural regions have been implicated in the processing of each of these, we propose to examine the neural correlates of functional categories in humans, at the whole brain level. For this purpose, we propose a new fMRI-based task inspired by the work of Pan et al. (2008, 2014) \cite{Pan2008-jm, Pan2014-lx}. Similar to the original monkey task, human participants would first, with reinforcement, learn to group unrelated visual stimuli into categories (as in Figure 3). Then, they would be tested to see if they can propagate reward associations from one stimulus to other members of the same category. Importantly, we can explicitly ask  human participants how confident they are that they will receive high/low reward when they are presented with a given stimulus, which provides us with a self-reported behavioral measure of generalization. Using reinforcement learning algorithms, we will model behavior during the learning phase(s) and use model parameters as predictors of future generalization performance. To arbitrate between the underlying mechanisms, we will investigate the representational content in specific brain areas previously linked with episodic inference and categorization (both perceptual and functional), examine the connectivity between these brain regions, as well as use model-based fMRI analyses to link trial-by-trial fluctuations in latent behavioral variables (estimated through computational modelling) to neural signals. More specifically, we will investigate whether the use of functional categories to make reward predictions is supported by separate or integrated memories in the striatum, LPFC, hippocampus, and/or VMPFC and examine how these regions work (or do not work) together. This work will shine a light on how humans can flexibly route relevant information across conceptual dimensions (i.e., between members of a functional category), one of the hallmarks of intelligence.		

\subsection{Question 3: What is the function of the LPFC in functional categorization?}
As mentioned above, functional categorization is related to episodic inference but distinct neural regions have been implicated for these via different tasks in distinct species: the hippocampus in episodic inference in humans and rodents and the LPFC in functional categorization in monkeys. While the proposed study in section 4.2. aims to resolve these discrepancies using human fMRI, the study proposed here will be a more direct follow up on the original work by Pan, Sakagami, and colleagues \cite{Pan2008-jm, Pan2014-lx}. It will use neural recording and chemo- and opto-genetic suppression techniques in monkeys, specifically in the hippocampus and the LPFC, to uncover the mechanisms of functional category formation. 

The transfer of information through arbitrary overlapping associations occurs in both episodic inference tasks (e.g. A- B, B-C, thus A -C) and the functional category task (e.g. because B1 and C1 are already known to belong to category 1, learning N1-B1 may cause processing of N1-category 1, category 1-large reward, thus N1-large reward). A key difference between them is that episodic inference tasks result in many unique groupings built on stimulus-stimulus associations (e.g. A1-B1-C1, A2-B2-C2, A3-B3-C3, A4-B4-C4, etc) but the functional category task results in just two groupings (e.g. high reward associated with all category 1 stimuli and low reward associated with all category 2 stimuli). Although the hippocampus is a key brain structure for the transfer of information across overlapping stimulus-stimulus associations in episodic inference tasks, its role in linking overlapping stimulus-stimulus associations into functional categories, so that information can be transferred between category members, remains to be determined. Are hippocampal stimulus-stimulus associations necessary for the formation of functional categories? And if so, are these alone sufficient or is LPFC involvement also necessary? (Because, in the extreme case, previous results showing functional category coding in the LPFC could simply reflect LPFC read-outs of hippocampal computations.)
			
Here, we propose a monkey experiment, inspired by the work of Pan, Sakagami, and colleagues \cite{Pan2008-jm, Pan2014-lx}, in which monkeys should first learn two categories of perceptually unrelated stimuli. For example, monkeys might learn that A1, A2, A3 belong in category 1 and that B1, B2, B3 belong in category 2. In the subsequent test period, novel stimuli will be associated with a member of each category. For example, monkeys might learn to select An over Bn after the presentations of A1 (A1→An). They will then be tested to see if they can make transitive inferences (e.g. that An should be selected over Bn after presentation of A2; A2→An), assumptions of symmetry (e.g. that A1 should be selected over B2 after presentation with An; An→A1), and transitive+symmetric inferences (e.g. that A2 should be selected over B2 after presentation with An; An→A2). 

In our proposed task, if neurons encoding a functional category can extend information from one member of the category to other members without direct experience, then they should reflect categorical information in tests involving transitivity and/or symmetry and the monkeys should have good behavioral performance. Whether or not hippocampal activity (representing stimulus-stimulus associations) is necessary for the formation of functional categories could be tested by chemo- or optogenetic suppression of the hippocampus during the learning of novel stimuli (An, Bn). If hippocampal activity is necessary then, after learning with a silenced hippocampus, neither behavioral responses nor neural activity in the hippocampus or LPFC would reflect category information. This should also lead to failures in transitive inference. 

If the hippocampus is found to be necessary for the formation of functional categories then the next question is whether or not this alone is sufficient. One alternative might be that the LPFC is also necessary. It is expected that the LPFC might be required for functional categorization, not only based on the work of Pan et al (2008, 2014), but also because it codes both sensory and motor information \cite{Hoshi2000-xl, Sakagami1999-rs}, even at a level of abstraction sufficient for the generalization of information (e.g. “motor category” neurons were found in the LPFC \cite{Shima2007-qu}). Hence, linking the sensory information of multiple stimulus-stimulus associations with motor information (shared response requirements between category members) may allow for formation of functional categories in the LPFC. 

If sensory-motor associations are indeed linked to form categories in the LPFC, then changing the required responses (e.g., putting the stimuli on the left/right instead of up/down positions of the screen, so that required responses would change from left/right to up/down) would induce systematic changes in the activity of category coding LPFC neurons. To test the necessity of the LPFC for functional categorization, in our proposed task chemo- or optogenetic suppression of the LPFC could be performed during the learning of novel stimuli (An, Bn). If LPFC activity is necessary (and so the hippocampus, without it, is not sufficient) then, after learning with a silenced LPFC, behavioral responses should not reflect category information. Note that this silencing of the LPFC during learning is also necessary to fairly examine hippocampal responses to see if they are sufficient for functional categorization. This is because the LPFC might otherwise convey category information to regions downstream, including possibly the hippocampus, as has previously been demonstrated with the inferior temporal cortex \cite{Tomita1999-ki}. If the hippocampus is not sufficient and the LPFC is indeed necessary then hippocampal neural responses to a given stimulus (e.g., A1), after learning with a silenced LPFC, should represent the paired specific stimulus (e.g., A2, A3, or An) and not its entire category (e.g. category 1). In this case, because the hippocampus would still be intact, transitive inference across stimulus-stimulus associations may still be found.

Examining these precise hypotheses should clarify the role of the hippocampus and the LPFC in functional categorization, as well as further clarifying the relation between episodic inference and functional categorization. Findings from this study will contribute a better integration of functional categorization within the broader field of generalization. 

\section{Final comments}
In this paper we highlighted the challenges in arbitrating between different theories of generalization. We further sought to find points of contact across the relevant research subfields. In short, we note that in the literature: the term “generalization” sometimes refers to a mechanism and sometimes to a behavioral outcome; what does and does not count as generalization may not be entirely agreed upon; research in different experimental species (humans, non-human primates, rodents) typically focuses on different questions; and a variety of generalization tasks are commonly used, with little clarity on whether these tap into the same or different processes. Overcoming these challenges will take time and it will take explicit effort to synthesize distinct lines of research. As a way forward in bridging some of the existing gaps in our understanding of generalization, we present three experimental proposals to examine questions that came up across the course of our GAC. Finally, we hope that this paper and new interactions across the broad fields of study within our GAC will help generate a fruitful framework for future generalization research. 	

\section*{acknowledgements}
This work was supported in part by National Institutes of Health [R01-NS112366 (DZ)], JST ERATO [JPMJER1801 (AC, JET)].

\section*{conflict of interest}
The authors declare no competing interests.

\printendnotes

\bibliography{references}

\begin{thebibliography}{163}
\providecommand{\natexlab}[1]{#1}
\providecommand{\url}[1]{\texttt{#1}}
\providecommand{\urlprefix}{}

\bibitem[{Jacobs and Liman(1991)Jacobs, Lucia F and Liman, Emily
  R}]{Jacobs1991-lj}
Jacobs LF, Liman ER.
\newblock Grey squirrels remember the locations of buried nuts.
\newblock Animal Behaviour 1991;41(1):103--110.

\bibitem[{McClelland et~al.(1995)McClelland, James L and McNaughton, Bruce L
  and O'Reilly, Randall C}]{McClelland1995-an}
McClelland JL, McNaughton BL, O'Reilly RC.
\newblock Why there are complementary learning systems in the hippocampus and
  neocortex: insights from the successes and failures of connectionist models
  of learning and memory.
\newblock Psychol Rev 1995 Jul;102(3):419--457.

\bibitem[{Poldrack and Packard(2003)Poldrack, Russell A and Packard, Mark
  G}]{Poldrack2003-uc}
Poldrack RA, Packard MG.
\newblock Competition among multiple memory systems: converging evidence from
  animal and human brain studies.
\newblock Neuropsychologia 2003;41(3):245--251.

\bibitem[{O'Reilly and Rudy(2000)O'Reilly, R C and Rudy, J W}]{OReilly2000-ti}
O'Reilly RC, Rudy JW.
\newblock Computational principles of learning in the neocortex and
  hippocampus.
\newblock Hippocampus 2000;10(4):389--397.

\bibitem[{Eichenbaum(1999)Eichenbaum, H}]{Eichenbaum1999-oa}
Eichenbaum H.
\newblock The hippocampus and mechanisms of declarative memory.
\newblock Behav Brain Res 1999 Sep;103(2):123--133.

\bibitem[{Zeithamova et~al.(2012)Zeithamova, Dagmar and Dominick, April L and
  Preston, Alison R}]{Zeithamova2012-ut}
Zeithamova D, Dominick AL, Preston AR.
\newblock Hippocampal and ventral medial prefrontal activation during
  retrieval-mediated learning supports novel inference.
\newblock Neuron 2012 Jul;75(1):168--179.

\bibitem[{Shohamy and Wagner(2008)Shohamy, Daphna and Wagner, Anthony
  D}]{Shohamy2008-qd}
Shohamy D, Wagner AD.
\newblock Integrating memories in the human brain: hippocampal-midbrain
  encoding of overlapping events.
\newblock Neuron 2008 Oct;60(2):378--389.

\bibitem[{Varga et~al.(2019)Varga, Nicole L and Gaugler, Trent and Talarico,
  Jennifer}]{Varga2019-bn}
Varga NL, Gaugler T, Talarico J.
\newblock Are mnemonic failures and benefits two sides of the same coin?:
  Investigating the real-world consequences of individual differences in memory
  integration.
\newblock Mem Cognit 2019 Apr;47(3):496--510.

\bibitem[{Knowlton and Squire(1993)Knowlton, B J and Squire, L
  R}]{Knowlton1993-xf}
Knowlton BJ, Squire LR.
\newblock The learning of categories: parallel brain systems for item memory
  and category knowledge.
\newblock Science 1993 Dec;262(5140):1747--1749.

\bibitem[{Squire and Knowlton(1995)Squire, L R and Knowlton, B
  J}]{Squire1995-uy}
Squire LR, Knowlton BJ.
\newblock Learning about categories in the absence of memory.
\newblock Proc Natl Acad Sci U S A 1995 Dec;92(26):12470--12474.

\bibitem[{Schlichting and Preston(2017)Schlichting, Margaret L and Preston,
  Alison R}]{Schlichting2017-aq}
Schlichting ML, Preston AR.
\newblock The Hippocampus and Memory Integration: Building Knowledge to
  Navigate Future Decisions.
\newblock In: Hannula DE, Duff MC, editors. The Hippocampus from Cells to
  Systems: Structure, Connectivity, and Functional Contributions to Memory and
  Flexible Cognition Cham: Springer International Publishing; 2017.p. 405--437.

\bibitem[{Zeithamova and Bowman(2020)Zeithamova, Dagmar and Bowman, Caitlin
  R}]{Zeithamova2020-sh}
Zeithamova D, Bowman CR.
\newblock Generalization and the hippocampus: More than one story?
\newblock Neurobiol Learn Mem 2020 Sep;p. 107317.

\bibitem[{Barron et~al.(2013)Barron, Helen C and Dolan, Raymond J and Behrens,
  Timothy E J}]{Barron2013-iy}
Barron HC, Dolan RJ, Behrens TEJ.
\newblock Online evaluation of novel choices by simultaneous representation of
  multiple memories.
\newblock Nat Neurosci 2013 Oct;16(10):1492--1498.

\bibitem[{Bowman and Zeithamova(2018)Bowman, Caitlin and Zeithamova,
  Dagmar}]{Bowman2018-pd}
Bowman C, Zeithamova D.
\newblock Abstract memory representations in the ventromedial prefrontal cortex
  and hippocampus support concept generalization.
\newblock J Neurosci 2018;38(10):2605--2614.

\bibitem[{Kumaran et~al.(2009)Kumaran, Dharshan and Summerfield, Jennifer J and
  Hassabis, Demis and Maguire, Eleanor A}]{Kumaran2009-cn}
Kumaran D, Summerfield JJ, Hassabis D, Maguire EA.
\newblock Tracking the emergence of conceptual knowledge during human decision
  making.
\newblock Neuron 2009 Sep;63(6):889--901.

\bibitem[{Schlichting et~al.(2015)Schlichting, Margaret L and Mumford, Jeanette
  A and Preston, Alison R}]{Schlichting2015-rj}
Schlichting ML, Mumford JA, Preston AR.
\newblock Learning-related representational changes reveal dissociable
  integration and separation signatures in the hippocampus and prefrontal
  cortex.
\newblock Nat Commun 2015 Aug;6:8151.

\bibitem[{Frank et~al.(2019)Frank, Lea E and Bowman, Caitlin R and Zeithamova,
  Dagmar}]{Frank2019-wx}
Frank LE, Bowman CR, Zeithamova D.
\newblock Differential Functional Connectivity along the Long Axis of the
  Hippocampus Aligns with Differential Role in Memory Specificity and
  Generalization.
\newblock J Cogn Neurosci 2019 Dec;31(12):1958--1975.

\bibitem[{Gerraty et~al.(2014)Gerraty, Raphael T and Davidow, Juliet Y and
  Wimmer, G Elliott and Kahn, Itamar and Shohamy, Daphna}]{Gerraty2014-ji}
Gerraty RT, Davidow JY, Wimmer GE, Kahn I, Shohamy D.
\newblock Transfer of learning relates to intrinsic connectivity between
  hippocampus, ventromedial prefrontal cortex, and large-scale networks.
\newblock J Neurosci 2014 Aug;34(34):11297--11303.

\bibitem[{van Kesteren et~al.(2010)van Kesteren, Marlieke T R and
  Fern{\'a}ndez, Guill{\'e}n and Norris, David G and Hermans, Erno
  J}]{Van_Kesteren2010-du}
van Kesteren MTR, Fern{\'a}ndez G, Norris DG, Hermans EJ.
\newblock Persistent schema-dependent hippocampal-neocortical connectivity
  during memory encoding and postencoding rest in humans.
\newblock Proc Natl Acad Sci U S A 2010 Apr;107(16):7550--7555.

\bibitem[{van Kesteren et~al.(2014)van Kesteren, Marlieke T R and Rijpkema,
  Mark and Ruiter, Dirk J and Morris, Richard G M and Fern{\'a}ndez,
  Guill{\'e}n}]{Van_Kesteren2014-ym}
van Kesteren MTR, Rijpkema M, Ruiter DJ, Morris RGM, Fern{\'a}ndez G.
\newblock Building on prior knowledge: schema-dependent encoding processes
  relate to academic performance.
\newblock J Cogn Neurosci 2014 Oct;26(10):2250--2261.

\bibitem[{Tolman(1948)Tolman, E C}]{Tolman1948-sn}
Tolman EC.
\newblock Cognitive maps in rats and men.
\newblock Psychol Rev 1948 Jul;55(4):189--208.

\bibitem[{O'Keefe and Dostrovsky(1971)O'Keefe, J and Dostrovsky,
  J}]{OKeefe1971-uy}
O'Keefe J, Dostrovsky J.
\newblock The hippocampus as a spatial map. Preliminary evidence from unit
  activity in the freely-moving rat.
\newblock Brain Res 1971;34:171--175.

\bibitem[{O'keefe and Nadel(1978)O'keefe, John and Nadel, Lynn}]{Okeefe1978-va}
O'keefe J, Nadel L.
\newblock The hippocampus as a cognitive map.
\newblock Oxford: Clarendon Press; 1978.

\bibitem[{Hafting et~al.(2005)Hafting, Torkel and Fyhn, Marianne and Molden,
  Sturla and Moser, May-Britt and Moser, Edvard}]{Hafting2005-jf}
Hafting T, Fyhn M, Molden S, Moser MB, Moser E.
\newblock Microstructure of a spatial map in the entorhinal cortex.
\newblock Nature 2005;436(7052):801--806.

\bibitem[{Hok et~al.(2005)Hok, V and Save, E and Lenck-Santini, P P and Poucet,
  B}]{Hok2005-yg}
Hok V, Save E, Lenck-Santini PP, Poucet B.
\newblock Coding for spatial goals in the prelimbic/infralimbic area of the rat
  frontal cortex.
\newblock Proc Natl Acad Sci U S A 2005 Mar;102(12):4602--4607.

\bibitem[{Doeller et~al.(2010)Doeller, Christian and Barry, Caswell and
  Burgess, Neil}]{Doeller2010-qv}
Doeller C, Barry C, Burgess N.
\newblock Evidence for grid cells in a human memory network.
\newblock Nature 2010;463(7281):657--661.

\bibitem[{Jacobs et~al.(2013)Jacobs, Joshua and Weidemann, Christoph and
  Miller, Jonathan and Solway, Alec and Burke, John and Wei, Xue-Xin and
  Suthana, Nanthia and Sperling, Michael and Sharan, Ashwini and Fried, Itzhak
  and Kahana, Michael}]{Jacobs2013-kb}
Jacobs J, Weidemann C, Miller J, Solway A, Burke J, Wei XX, et~al.
\newblock Direct recordings of grid-like neuronal activity in human spatial
  navigation.
\newblock Nat Neurosci 2013;16(9):1188--1190.

\bibitem[{Yartsev and Ulanovsky(2013)Yartsev, Michael M and Ulanovsky,
  Nachum}]{Yartsev2013-dg}
Yartsev MM, Ulanovsky N.
\newblock Representation of three-dimensional space in the hippocampus of
  flying bats.
\newblock Science 2013 Apr;340(6130):367--372.

\bibitem[{Constantinescu et~al.(2016)Constantinescu, Alexandra O and O'Reilly,
  Jill X and Behrens, Timothy E J}]{Constantinescu2016-ys}
Constantinescu AO, O'Reilly JX, Behrens TEJ.
\newblock Organizing conceptual knowledge in humans with a gridlike code.
\newblock Science 2016 Jun;352(6292):1464--1468.

\bibitem[{Bellmund et~al.(2018)Bellmund, Jacob L S and G{\"a}rdenfors, Peter
  and Moser, Edvard I and Doeller, Christian F}]{Bellmund2018-vg}
Bellmund JLS, G{\"a}rdenfors P, Moser EI, Doeller CF.
\newblock Navigating cognition: Spatial codes for human thinking.
\newblock Science 2018 Nov;362(6415).

\bibitem[{Mok and Love(2019)Mok, Robert M and Love, Bradley C}]{Mok2019-zg}
Mok RM, Love BC.
\newblock A non-spatial account of place and grid cells based on clustering
  models of concept learning.
\newblock Nat Commun 2019 Dec;10(1):5685.

\bibitem[{Stachenfeld et~al.(2017)Stachenfeld, Kimberly and Botvinick, Matthew
  and Gershman, Samuel}]{Stachenfeld2017-tj}
Stachenfeld K, Botvinick M, Gershman S.
\newblock The hippocampus as a predictive map.
\newblock Nat Neurosci 2017;.

\bibitem[{Whittington et~al.(2020)Whittington, James C R and Muller, Timothy H
  and Mark, Shirley and Chen, Guifen and Barry, Caswell and Burgess, Neil and
  Behrens, Timothy E J}]{Whittington2020-uq}
Whittington JCR, Muller TH, Mark S, Chen G, Barry C, Burgess N, et~al.
\newblock The {Tolman-Eichenbaum} Machine: Unifying Space and Relational Memory
  through Generalization in the Hippocampal Formation.
\newblock Cell 2020 Nov;0(0).

\bibitem[{Carpenter et~al.(2015)Carpenter, Francis and Manson, Daniel and
  Jeffery, Kate and Burgess, Neil and Barry, Caswell}]{Carpenter2015-vj}
Carpenter F, Manson D, Jeffery K, Burgess N, Barry C.
\newblock Grid cells form a global representation of connected environments.
\newblock Curr Biol 2015 May;25(9):1176--1182.

\bibitem[{Bellmund et~al.(2016)Bellmund, Jacob Ls and Deuker, Lorena and
  Navarro Schr{\"o}der, Tobias and Doeller, Christian F}]{Bellmund2016-yz}
Bellmund JL, Deuker L, Navarro~Schr{\"o}der T, Doeller CF.
\newblock Grid-cell representations in mental simulation.
\newblock Elife 2016 Aug;5.

\bibitem[{Vigan{\`o} et~al.(2021)Vigan{\`o}, Simone and Borghesani, Valentina
  and Piazza, Manuela}]{Vigano2021-gz}
Vigan{\`o} S, Borghesani V, Piazza M.
\newblock Symbolic categorization of novel multisensory stimuli in the human
  brain.
\newblock Neuroimage 2021 Apr;p. 118016.

\bibitem[{Park et~al.(2020)Park, Seongmin A and Miller, Douglas S and Nili,
  Hamed and Ranganath, Charan and Boorman, Erie D}]{Park2020-ng}
Park SA, Miller DS, Nili H, Ranganath C, Boorman ED.
\newblock Map Making: Constructing, Combining, and Inferring on Abstract
  Cognitive Maps.
\newblock Neuron 2020 Jul;0(0).

\bibitem[{Curtis and Jamieson(2019)Curtis, Evan T and Jamieson, Randall
  K}]{Curtis2019-ww}
Curtis ET, Jamieson RK.
\newblock Computational and empirical simulations of selective memory
  impairments: Converging evidence for a single-system account of memory
  dissociations.
\newblock Q J Exp Psychol 2019 Apr;72(4):798--817.

\bibitem[{Hintzman and Ludlam(1980)Hintzman, D L and Ludlam,
  G}]{Hintzman1980-ms}
Hintzman DL, Ludlam G.
\newblock Differential forgetting of prototypes and old instances: simulation
  by an exemplar-based classification model.
\newblock Mem Cognit 1980 Jul;8(4):378--382.

\bibitem[{Kinder and Shanks(2001)Kinder, A and Shanks, D R}]{Kinder2001-cb}
Kinder A, Shanks DR.
\newblock Amnesia and the declarative/nondeclarative distinction: a recurrent
  network model of classification, recognition, and repetition priming.
\newblock J Cogn Neurosci 2001 Jul;13(5):648--669.

\bibitem[{Zaki et~al.(2003)Zaki, Safa R and Nosofsky, Robert M and Jessup,
  Nenette M and Unverzagt, Frederick W}]{Zaki2003-oq}
Zaki SR, Nosofsky RM, Jessup NM, Unverzagt FW.
\newblock Categorization and recognition performance of a memory-impaired
  group: evidence for single-system models.
\newblock J Int Neuropsychol Soc 2003 Mar;9(3):394--406.

\bibitem[{Kumaran and McClelland(2012)Kumaran, Dharshan and McClelland, James
  L}]{Kumaran2012-gs}
Kumaran D, McClelland JL.
\newblock Generalization through the recurrent interaction of episodic
  memories: a model of the hippocampal system.
\newblock Psychol Rev 2012 Jul;119(3):573--616.

\bibitem[{Carpenter and Schacter(2017)Carpenter, Alexis C and Schacter, Daniel
  L}]{Carpenter2017-bi}
Carpenter AC, Schacter DL.
\newblock Flexible retrieval: When true inferences produce false memories.
\newblock J Exp Psychol Learn Mem Cogn 2017 Mar;43(3):335--349.

\bibitem[{Koster et~al.(2018)Koster, Raphael and Chadwick, Martin J and Chen,
  Yi and Berron, David and Banino, Andrea and D{\"u}zel, Emrah and Hassabis,
  Demis and Kumaran, Dharshan}]{Koster2018-sb}
Koster R, Chadwick MJ, Chen Y, Berron D, Banino A, D{\"u}zel E, et~al.
\newblock {Big-Loop} Recurrence within the Hippocampal System Supports
  Integration of Information across Episodes.
\newblock Neuron 2018 Sep;99(6):1342--1354.e6.

\bibitem[{Barron et~al.(2020)Barron, Helen C and Reeve, Hayley M and
  Koolschijn, Ren{\'e}e S and Perestenko, Pavel V and Shpektor, Anna and Nili,
  Hamed and Rothaermel, Roman and Campo-Urriza, Natalia and O'Reilly, Jill X
  and Bannerman, David M and Behrens, Timothy E J and Dupret,
  David}]{Barron2020-bm}
Barron HC, Reeve HM, Koolschijn RS, Perestenko PV, Shpektor A, Nili H, et~al.
\newblock Neuronal Computation Underlying Inferential Reasoning in Humans and
  Mice.
\newblock Cell 2020 Oct;183(1):228--243.e21.

\bibitem[{Wiltgen and Silva(2007)Wiltgen, Brian J and Silva, Alcino
  J}]{Wiltgen2007-jq}
Wiltgen BJ, Silva AJ.
\newblock Memory for context becomes less specific with time.
\newblock Learn Mem 2007 Apr;14(4):313--317.

\bibitem[{Diekelmann and Born(2010)Diekelmann, Susanne and Born,
  Jan}]{Diekelmann2010-wy}
Diekelmann S, Born J.
\newblock The memory function of sleep.
\newblock Nat Rev Neurosci 2010 Feb;11(2):114--126.

\bibitem[{Ellenbogen et~al.(2007)Ellenbogen, Jeffrey M and Hu, Peter T and
  Payne, Jessica D and Titone, Debra and Walker, Matthew P}]{Ellenbogen2007-se}
Ellenbogen JM, Hu PT, Payne JD, Titone D, Walker MP.
\newblock Human relational memory requires time and sleep.
\newblock Proc Natl Acad Sci U S A 2007 May;104(18):7723--7728.

\bibitem[{Lau et~al.(2011)Lau, Hiuyan and Alger, Sara E and Fishbein,
  William}]{Lau2011-hy}
Lau H, Alger SE, Fishbein W.
\newblock Relational memory: a daytime nap facilitates the abstraction of
  general concepts.
\newblock PLoS One 2011 Nov;6(11):e27139.

\bibitem[{Schapiro et~al.(2017)Schapiro, Anna C and McDevitt, Elizabeth A and
  Chen, Lang and Norman, Kenneth A and Mednick, Sara C and Rogers, Timothy
  T}]{Schapiro2017-tw}
Schapiro AC, McDevitt EA, Chen L, Norman KA, Mednick SC, Rogers TT.
\newblock Sleep Benefits Memory for Semantic Category Structure While
  Preserving {Exemplar-Specific} Information.
\newblock Sci Rep 2017 Nov;7(1):14869.

\bibitem[{Tompary and Davachi(2017)Tompary, Alexa and Davachi,
  Lila}]{Tompary2017-gd}
Tompary A, Davachi L.
\newblock Consolidation Promotes the Emergence of Representational Overlap in
  the Hippocampus and Medial Prefrontal Cortex.
\newblock Neuron 2017 Sep;96(1):228--241.e5.

\bibitem[{Buzs{\'a}ki(2015)Buzs{\'a}ki, Gy{\"o}rgy}]{Buzsaki2015-qh}
Buzs{\'a}ki G.
\newblock Hippocampal sharp wave-ripple: A cognitive biomarker for episodic
  memory and planning.
\newblock Hippocampus 2015 Oct;25(10):1073--1188.

\bibitem[{Joo and Frank(2018)Joo, Hannah R and Frank, Loren M}]{Joo2018-ho}
Joo HR, Frank LM.
\newblock The hippocampal sharp wave-ripple in memory retrieval for immediate
  use and consolidation.
\newblock Nat Rev Neurosci 2018 Dec;19(12):744--757.

\bibitem[{Foster(2017)Foster, David J}]{Foster2017-xw}
Foster DJ.
\newblock Replay Comes of Age.
\newblock Annu Rev Neurosci 2017 Jul;40:581--602.

\bibitem[{Dragoi and Tonegawa(2011)Dragoi, George and Tonegawa,
  Susumu}]{Dragoi2011-qp}
Dragoi G, Tonegawa S.
\newblock Preplay of future place cell sequences by hippocampal cellular
  assemblies.
\newblock Nature 2011 Jan;469(7330):397--401.

\bibitem[{{\'O}lafsd{\'o}ttir et~al.(2015){\'O}lafsd{\'o}ttir, H Freyja and
  Barry, Caswell and Saleem, Aman B and Hassabis, Demis and Spiers, Hugo
  J}]{Olafsdottir2015-wp}
{\'O}lafsd{\'o}ttir HF, Barry C, Saleem AB, Hassabis D, Spiers HJ.
\newblock Hippocampal place cells construct reward related sequences through
  unexplored space.
\newblock Elife 2015 Jun;4:e06063.

\bibitem[{Liu et~al.(2019)Liu, Yunzhe and Dolan, Raymond J and Kurth-Nelson,
  Zeb and Behrens, Timothy E J}]{Liu2019-cu}
Liu Y, Dolan RJ, Kurth-Nelson Z, Behrens TEJ.
\newblock Human Replay Spontaneously Reorganizes Experience.
\newblock Cell 2019 Jul;.

\bibitem[{Ji and Wilson(2007)Ji, Daoyun and Wilson, Matthew A}]{Ji2007-dn}
Ji D, Wilson MA.
\newblock Coordinated memory replay in the visual cortex and hippocampus during
  sleep.
\newblock Nat Neurosci 2007 Jan;10(1):100--107.

\bibitem[{Qin et~al.(1997)Qin, Y L and McNaughton, B L and Skaggs, W E and
  Barnes, C A}]{Qin1997-jl}
Qin YL, McNaughton BL, Skaggs WE, Barnes CA.
\newblock Memory reprocessing in corticocortical and hippocampocortical
  neuronal ensembles.
\newblock Philos Trans R Soc Lond B Biol Sci 1997 Oct;352(1360):1525--1533.

\bibitem[{Ashby(1992)Ashby, F Gregory}]{Ashby1992-pl}
Ashby FG.
\newblock Multidimensional models of categorization.
\newblock Multidimensional models of perception and cognition
  1992;523:449--483.

\bibitem[{Ashby and Maddox(1993)Ashby, F Gregory and Maddox, W
  Todd}]{Ashby1993-bi}
Ashby FG, Maddox WT.
\newblock Relations between Prototype, Exemplar, and Decision Bound Models of
  Categorization.
\newblock J Math Psychol 1993 Sep;37(3):372--400.

\bibitem[{Maddox and Ashby(1993)Maddox, W T and Ashby, F G}]{Maddox1993-oh}
Maddox WT, Ashby FG.
\newblock Comparing decision bound and exemplar models of categorization.
\newblock Percept Psychophys 1993 Jan;53(1):49--70.

\bibitem[{Ashby et~al.(1998)Ashby, F G and Alfonso-Reese, L A and Turken, A U
  and Waldron, E M}]{Ashby1998-uy}
Ashby FG, Alfonso-Reese LA, Turken AU, Waldron EM.
\newblock A neuropsychological theory of multiple systems in category learning.
\newblock Psychol Rev 1998 Jul;105(3):442--481.

\bibitem[{Gregory~Ashby and Wang(2004)Gregory Ashby, F and Wang,
  Yi-Wen}]{Gregory_Ashby_undated-rh}
Gregory~Ashby F, Wang YW.
\newblock Computational Neuroscientific Models of Categorization.
\newblock In: Sun R, editor. The Cambridge Handbook of Computational Cognitive
  Sciences New York: Cambridge University Press; 2004.p. tbd.

\bibitem[{Nomura et~al.(2006)Nomura, E M and Maddox, W T and Filoteo, J V and
  Ing, A D and Gitelman, D R and Parrish, T B and Mesulam, M-M and Reber, P
  J}]{Nomura2006-av}
Nomura EM, Maddox WT, Filoteo JV, Ing AD, Gitelman DR, Parrish TB, et~al.
\newblock Neural Correlates of {Rule-Based} and {Information-Integration}
  Visual Category Learning.
\newblock Cereb Cortex 2006 Jan;17(1):37--43.

\bibitem[{Nosofsky and Johansen(2000)Nosofsky, Robert M and Johansen, Mark
  K}]{Nosofsky2000-fl}
Nosofsky RM, Johansen MK.
\newblock Exemplar-based accounts of ``multiple-system'' phenomena in
  perceptual categorization.
\newblock Psychon Bull Rev 2000 Sep;7(3):375--402.

\bibitem[{Pothos(2005)Pothos, Emmanuel M}]{Pothos2005-yt}
Pothos EM.
\newblock The rules versus similarity distinction.
\newblock Behav Brain Sci 2005 Feb;28(1):1--14; discussion 14--49.

\bibitem[{Verguts and Fias(2009)Verguts, Tom and Fias, Wim}]{Verguts2009-vw}
Verguts T, Fias W.
\newblock Similarity and rules United: similarity- and rule-based processing in
  a single neural network.
\newblock Cogn Sci 2009 Mar;33(2):243--259.

\bibitem[{Maddox and Ashby(2004)Maddox, W Todd and Ashby, F
  Gregory}]{Maddox2004-kc}
Maddox WT, Ashby FG.
\newblock Dissociating explicit and procedural-learning based systems of
  perceptual category learning.
\newblock Behav Processes 2004 Jun;66(3):309--332.

\bibitem[{Casale et~al.(2012)Casale, Michael B and Roeder, Jessica L and Ashby,
  F Gregory}]{Casale2012-dx}
Casale MB, Roeder JL, Ashby FG.
\newblock Analogical transfer in perceptual categorization.
\newblock Mem Cognit 2012 Apr;40(3):434--449.

\bibitem[{Smith et~al.(2015)Smith, J David and Zakrzewski, Alexandria C and
  Johnston, Jennifer J R and Roeder, Jessica L and Boomer, Joseph and Ashby, F
  Gregory and Church, Barbara A}]{Smith2015-hz}
Smith JD, Zakrzewski AC, Johnston JJR, Roeder JL, Boomer J, Ashby FG, et~al.
\newblock Generalization of category knowledge and dimensional categorization
  in humans (Homo sapiens) and nonhuman primates (Macaca mulatta).
\newblock J Exp Psychol Anim Learn Cogn 2015 Oct;41(4):322--335.

\bibitem[{Qadri et~al.(2019)Qadri, Muhammad A J and Ashby, F Gregory and Smith,
  J David and Cook, Robert G}]{Qadri2019-ib}
Qadri MAJ, Ashby FG, Smith JD, Cook RG.
\newblock Testing analogical rule transfer in pigeons (Columba livia).
\newblock Cognition 2019 Feb;183:256--268.

\bibitem[{Till and Priluck(2000)Till, Brian D and Priluck, Randi
  Lynn}]{Till2000-vz}
Till BD, Priluck RL.
\newblock Stimulus generalization in classical conditioning: An initial
  investigation and extension.
\newblock Psychol Mark 2000 Jan;17(1):55--72.

\bibitem[{Brogden(1939)Brogden, W J}]{Brogden1939-kd}
Brogden WJ.
\newblock Sensory pre-conditioning.
\newblock J Exp Psychol 1939;25(4):323--332.

\bibitem[{Yakovlev et~al.(1998)Yakovlev, V and Fusi, S and Berman, E and
  Zohary, E}]{Yakovlev1998-qd}
Yakovlev V, Fusi S, Berman E, Zohary E.
\newblock Inter-trial neuronal activity in inferior temporal cortex: a putative
  vehicle to generate long-term visual associations.
\newblock Nat Neurosci 1998 Aug;1(4):310--317.

\bibitem[{Nosofsky(1986)Nosofsky, Robert M}]{Nosofsky1986-pb}
Nosofsky RM.
\newblock Attention, similarity, and the identification--categorization
  relationship.
\newblock J Exp Psychol Gen 1986 Mar;115(1):39--57.

\bibitem[{.~Rips(1989). Rips, Lance J}]{Rips1989-yd}
~Rips LJ.
\newblock Similarity, typicality, and categorization.
\newblock In: Similarity and Analogical Reasoning Cambridge University Press;
  1989.p. 21--59.

\bibitem[{Shanks and Darby(1998)Shanks, David R and Darby, Richard
  J}]{Shanks1998-xs}
Shanks DR, Darby RJ.
\newblock Feature- and rule-based generalization in human associative learning.
\newblock J Exp Psychol Anim Behav Process 1998;24(4):405--415.

\bibitem[{Zeithamova and Preston(2010)Zeithamova, Dagmar and Preston, Alison
  R}]{Zeithamova2010-ux}
Zeithamova D, Preston AR.
\newblock Flexible memories: differential roles for medial temporal lobe and
  prefrontal cortex in cross-episode binding.
\newblock J Neurosci 2010 Nov;30(44):14676--14684.

\bibitem[{Lewis and Durrant(2011)Lewis, Penelope A and Durrant, Simon
  J}]{Lewis2011-ih}
Lewis PA, Durrant SJ.
\newblock Overlapping memory replay during sleep builds cognitive schemata.
\newblock Trends Cogn Sci 2011 Aug;15(8):343--351.

\bibitem[{Witkowski et~al.(2020)Witkowski, Sarah and Schechtman, Eitan and
  Paller, Ken A}]{Witkowski2020-dh}
Witkowski S, Schechtman E, Paller KA.
\newblock Examining sleep's role in memory generalization and specificity
  through the lens of targeted memory reactivation.
\newblock Current Opinion in Behavioral Sciences 2020 Jun;33:86--91.

\bibitem[{Recht et~al.(2019)Recht, Benjamin and Roelofs, Rebecca and Schmidt,
  Ludwig and Shankar, Vaishaal}]{Recht2019-gi}
Recht B, Roelofs R, Schmidt L, Shankar V.
\newblock Do Image Net Classifiers Generalize to Image Net?
\newblock In: Proceedings of the 36th International Conference on Machine
  Learning; 2019. p. 5389--5400.

\bibitem[{Lu et~al.(2019)Lu, Alex X and Lu, Amy X and Schormann, Wiebke and
  Ghassemi, Marzyeh and Andrews, David W and Moses, Alan M}]{Lu2019-cf}
Lu AX, Lu AX, Schormann W, Ghassemi M, Andrews DW, Moses AM.
\newblock The Cells Out of Sample ({COOS}) dataset and benchmarks for measuring
  out-of-sample generalization of image classifiers; 2019, -.

\bibitem[{Atkeson et~al.(2018)Atkeson, Christopher G and Benzun, P W Babu and
  Banerjee, Nandan and Berenson, Dmitry and Bove, Christoper P and Cui, Xiongyi
  and DeDonato, Mathew and Du, Ruixiang and Feng, Siyuan and Franklin, Perry
  and Gennert, M and Graff, Joshua P and He, Peng and Jaeger, Aaron and Kim,
  Joohyung and Knoedler, Kevin and Li, Lening and Liu, Chenggang and Long,
  Xianchao and Padir, T and Polido, Felipe and Tighe, G G and Xinjilefu,
  X}]{Atkeson2018-du}
Atkeson CG, Benzun PWB, Banerjee N, Berenson D, Bove CP, Cui X, et~al.
\newblock What Happened at the {DARPA} Robotics Challenge Finals.
\newblock In: Spenko M, Buerger S, Iagnemma K, editors. The {DARPA} Robotics
  Challenge Finals: Humanoid Robots To The Rescue Cham: Springer International
  Publishing; 2018.p. 667--684.

\bibitem[{Shepard et~al.(1961)Shepard, Roger N and Hovland, Carl I and Jenkins,
  Herbert M}]{Shepard1961-ll}
Shepard RN, Hovland CI, Jenkins HM.
\newblock Learning and memorization of classifications.
\newblock Psychol Monogr 1961;75(13):1--42.

\bibitem[{Wasserman et~al.(1992)Wasserman, E A and DeVolder, C L and Coppage, D
  J}]{Wasserman1992-ym}
Wasserman EA, DeVolder CL, Coppage DJ.
\newblock Non-Similarity-Based Conceptualization in Pigeons via Secondary or
  Mediated Generalization.
\newblock Psychological Science 1992;3(6):374--379.

\bibitem[{Hull(1943)Hull, C L}]{Hull1943-ts}
Hull CL.
\newblock Principles of behavior: An introduction to behavior theory.
\newblock Century psychology series, New York : Appleton-Century Company,
  Incorporated; 1943.

\bibitem[{Bialer(1961)Bialer, I}]{Bialer1961-ca}
Bialer I.
\newblock Primary and secondary stimulus generalization as related to
  intelligence level.
\newblock J Exp Psychol 1961 Oct;62:395--402.

\bibitem[{Davis et~al.(2017)Davis, Tyler and Goldwater, Micah and Giron,
  Josue}]{Davis2017-fy}
Davis T, Goldwater M, Giron J.
\newblock From Concrete Examples to Abstract Relations: The Rostrolateral
  Prefrontal Cortex Integrates Novel Examples into Relational Categories.
\newblock Cereb Cortex 2017 Apr;27(4):2652--2670.

\bibitem[{Wutz et~al.(2018)Wutz, A and Loonis, R and Roy, J and Donoghue, J and
  Miller, E}]{Wutz2018-wm}
Wutz A, Loonis R, Roy J, Donoghue J, Miller E.
\newblock Different Levels of Category Abstraction by Different Dynamics in
  Different Prefrontal Areas.
\newblock Neuron 2018;97(3):716--726.

\bibitem[{Dusek and Eichenbaum(1997)Dusek, J A and Eichenbaum,
  H}]{Dusek1997-hf}
Dusek JA, Eichenbaum H.
\newblock The hippocampus and memory for orderly stimulus relations.
\newblock Proc Natl Acad Sci U S A 1997 Jun;94(13):7109--7114.

\bibitem[{Preston et~al.(2004)Preston, Alison R and Shrager, Yael and
  Dudukovic, Nicole M and Gabrieli, John D E}]{Preston2004-tq}
Preston AR, Shrager Y, Dudukovic NM, Gabrieli JDE.
\newblock Hippocampal contribution to the novel use of relational information
  in declarative memory.
\newblock Hippocampus 2004;14(2):148--152.

\bibitem[{Iordanova et~al.(2011)Iordanova, Mihaela D and Good, Mark and Honey,
  Robert C}]{Iordanova2011-ex}
Iordanova MD, Good M, Honey RC.
\newblock Retrieval-mediated learning involving episodes requires synaptic
  plasticity in the hippocampus.
\newblock J Neurosci 2011 May;31(19):7156--7162.

\bibitem[{Mazur(2015)Mazur, James E}]{Mazur2015-pg}
Mazur JE.
\newblock Learning and behavior: Instructor's review copy.
\newblock Psychology Press; 2015.

\bibitem[{Honey and Hall(1989)Honey, R C and Hall, G}]{Honey1989-lf}
Honey RC, Hall G.
\newblock Acquired equivalence and distinctiveness of cues.
\newblock J Exp Psychol Anim Behav Process 1989 Oct;15(4):338--346.

\bibitem[{Honey(1990)Honey, R C}]{Honey1990-xl}
Honey RC.
\newblock Stimulus generalization as a function of stimulus novelty and
  familiarity in rats.
\newblock J Exp Psychol Anim Behav Process 1990 Apr;16(2):178--184.

\bibitem[{Maes et~al.(2015)Maes, Elisa and De Filippo, Guido and Inkster, Angus
  B and Lea, Stephen E G and De Houwer, Jan and D'Hooge, Rudi and Beckers, Tom
  and Wills, Andy J}]{Maes2015-yl}
Maes E, De~Filippo G, Inkster AB, Lea SEG, De~Houwer J, D'Hooge R, et~al.
\newblock Feature- versus rule-based generalization in rats, pigeons and
  humans.
\newblock Anim Cogn 2015 Nov;18(6):1267--1284.

\bibitem[{Murphy et~al.(2008)Murphy, Robin A and Mondrag{\'o}n, Esther and
  Murphy, Victoria A}]{Murphy2008-uw}
Murphy RA, Mondrag{\'o}n E, Murphy VA.
\newblock Rule learning by rats.
\newblock Science 2008 Mar;319(5871):1849--1851.

\bibitem[{Pan et~al.(2008)Pan, Xiaochuan and Sawa, Kosuke and Tsuda, Ichiro and
  Tsukada, Minoru and Sakagami, Masamichi}]{Pan2008-jm}
Pan X, Sawa K, Tsuda I, Tsukada M, Sakagami M.
\newblock Reward prediction based on stimulus categorization in primate lateral
  prefrontal cortex.
\newblock Nat Neurosci 2008;11(6):703--712.

\bibitem[{Pan et~al.(2014)Pan, Xiaochuan and Fan, Hongwei and Sawa, Kosuke and
  Tsuda, Ichiro and Tsukada, Minoru and Sakagami, Masamichi}]{Pan2014-lx}
Pan X, Fan H, Sawa K, Tsuda I, Tsukada M, Sakagami M.
\newblock Reward inference by primate prefrontal and striatal neurons.
\newblock J Neurosci 2014 Jan;34(4):1380--1396.

\bibitem[{Bowman et~al.(2020)Bowman, Caitlin R and Iwashita, Takako and
  Zeithamova, Dasa}]{Bowman2020-hv}
Bowman CR, Iwashita T, Zeithamova D.
\newblock Tracking prototype and exemplar representations in the brain across
  learning.
\newblock Elife 2020 Nov;9.

\bibitem[{Schlichting and Preston(2016)Schlichting, Margaret L and Preston,
  Alison R}]{Schlichting2016-co}
Schlichting ML, Preston AR.
\newblock Hippocampal-medial prefrontal circuit supports memory updating during
  learning and post-encoding rest.
\newblock Neurobiol Learn Mem 2016 Oct;134 Pt A:91--106.

\bibitem[{van Kesteren et~al.(2013)van Kesteren, Marlieke T R and Beul, Sarah F
  and Takashima, Atsuko and Henson, Richard N and Ruiter, Dirk J and
  Fern{\'a}ndez, Guill{\'e}n}]{Van_Kesteren2013-yh}
van Kesteren MTR, Beul SF, Takashima A, Henson RN, Ruiter DJ, Fern{\'a}ndez G.
\newblock Differential roles for medial prefrontal and medial temporal cortices
  in schema-dependent encoding: from congruent to incongruent.
\newblock Neuropsychologia 2013 Oct;51(12):2352--2359.

\bibitem[{Warren et~al.(2014)Warren, David E and Jones, Samuel H and Duff,
  Melissa C and Tranel, Daniel}]{Warren2014-gw}
Warren DE, Jones SH, Duff MC, Tranel D.
\newblock False recall is reduced by damage to the ventromedial prefrontal
  cortex: implications for understanding the neural correlates of schematic
  memory.
\newblock J Neurosci 2014 May;34(22):7677--7682.

\bibitem[{DeVito et~al.(2010)DeVito, Loren M and Lykken, Christine and Kanter,
  Benjamin R and Eichenbaum, Howard}]{DeVito2010-tf}
DeVito LM, Lykken C, Kanter BR, Eichenbaum H.
\newblock Prefrontal cortex: role in acquisition of overlapping associations
  and transitive inference.
\newblock Learn Mem 2010 Mar;17(3):161--167.

\bibitem[{Freedman et~al.(2001)Freedman, D J and Riesenhuber, M and Poggio, T
  and Miller, E K}]{Freedman2001-zu}
Freedman DJ, Riesenhuber M, Poggio T, Miller EK.
\newblock Categorical representation of visual stimuli in the primate
  prefrontal cortex.
\newblock Science 2001 Jan;291(5502):312--316.

\bibitem[{Neubert et~al.(2014)Neubert, Franz-Xaver and Mars, Rogier and Thomas,
  Adam and Sallet, Jerome and Rushworth, Matthew}]{Neubert2014-fu}
Neubert FX, Mars R, Thomas A, Sallet J, Rushworth M.
\newblock Comparison of Human Ventral Frontal Cortex Areas for Cognitive
  Control and Language with Areas in Monkey Frontal Cortex.
\newblock Neuron 2014;81(3):700--713.

\bibitem[{Neubert et~al.(2015)Neubert, Franz-Xaver and Mars, Rogier B and
  Sallet, J{\'e}r{\^o}me and Rushworth, Matthew F S}]{Neubert2015-ja}
Neubert FX, Mars RB, Sallet J, Rushworth MFS.
\newblock Connectivity reveals relationship of brain areas for reward-guided
  learning and decision making in human and monkey frontal cortex.
\newblock Proc Natl Acad Sci U S A 2015 May;112(20):E2695--704.

\bibitem[{Heckers et~al.(2004)Heckers, Stephan and Zalesak, Martin and Weiss,
  Anthony P and Ditman, Tali and Titone, Debra}]{Heckers2004-ri}
Heckers S, Zalesak M, Weiss AP, Ditman T, Titone D.
\newblock Hippocampal activation during transitive inference in humans.
\newblock Hippocampus 2004;14(2):153--162.

\bibitem[{Ryan et~al.(2016)Ryan, Jennifer D and D'Angelo, Maria C and Kamino,
  Daphne and Ostreicher, Melanie and Moses, Sandra N and Rosenbaum, R
  Shayna}]{Ryan2016-xk}
Ryan JD, D'Angelo MC, Kamino D, Ostreicher M, Moses SN, Rosenbaum RS.
\newblock Relational learning and transitive expression in aging and amnesia.
\newblock Hippocampus 2016 Feb;26(2):170--184.

\bibitem[{Zalesak and Heckers(2009)Zalesak, Martin and Heckers,
  Stephan}]{Zalesak2009-yf}
Zalesak M, Heckers S.
\newblock The role of the hippocampus in transitive inference.
\newblock Psychiatry Res 2009 Apr;172(1):24--30.

\bibitem[{Preston and Eichenbaum(2013)Preston, Alison R and Eichenbaum,
  Howard}]{Preston2013-kd}
Preston AR, Eichenbaum H.
\newblock Interplay of hippocampus and prefrontal cortex in memory.
\newblock Curr Biol 2013 Sep;23(17):R764--73.

\bibitem[{Bunsey and Eichenbaum(1996)Bunsey, M and Eichenbaum,
  H}]{Bunsey1996-gm}
Bunsey M, Eichenbaum H.
\newblock Conservation of hippocampal memory function in rats and humans.
\newblock Nature 1996 Jan;379(6562):255--257.

\bibitem[{Posner and Keele(1968)Posner, M I and Keele, S W}]{Posner1968-hn}
Posner MI, Keele SW.
\newblock On the genesis of abstract ideas.
\newblock J Exp Psychol 1968 Jul;77(3):353--363.

\bibitem[{Zeithamova and Maddox(2006)Zeithamova, Dagmar and Maddox, W
  Todd}]{Zeithamova2006-en}
Zeithamova D, Maddox WT.
\newblock Dual-task interference in perceptual category learning.
\newblock Mem Cognit 2006 Mar;34(2):387--398.

\bibitem[{Davis et~al.(2012)Davis, Tyler and Love, Bradley C and Preston,
  Alison R}]{Davis2012-ke}
Davis T, Love BC, Preston AR.
\newblock Learning the exception to the rule: model-based {FMRI} reveals
  specialized representations for surprising category members.
\newblock Cereb Cortex 2012 Feb;22(2):260--273.

\bibitem[{Bozoki et~al.(2006)Bozoki, Andrea and Grossman, Murray and Smith,
  Edward E}]{Bozoki2006-ej}
Bozoki A, Grossman M, Smith EE.
\newblock Can patients with Alzheimer's disease learn a category implicitly?
\newblock Neuropsychologia 2006;44(5):816--827.

\bibitem[{Watanabe et~al.(1995)Watanabe, S and Sakamoto, J and Wakita,
  M}]{Watanabe1995-jo}
Watanabe S, Sakamoto J, Wakita M.
\newblock Pigeons' discrimination of paintings by Monet and Picasso.
\newblock J Exp Anal Behav 1995 Mar;63(2):165--174.

\bibitem[{Schapiro et~al.(2017)Schapiro, Anna and Turk-Browne, Nicholas and
  Botvinick, Matthew and Norman, Kenneth}]{Schapiro2017-nb}
Schapiro A, Turk-Browne N, Botvinick M, Norman K.
\newblock Complementary learning systems within the hippocampus: a neural
  network modelling approach to reconciling episodic memory with statistical
  learning.
\newblock Philos Trans R Soc Lond B Biol Sci 2017;372(1711):20160049.

\bibitem[{Ashby et~al.(2020)Ashby, Stefania R and Bowman, Caitlin R and
  Zeithamova, Dagmar}]{Ashby2020-ok}
Ashby SR, Bowman CR, Zeithamova D.
\newblock Perceived similarity ratings predict generalization success after
  traditional category learning and a new paired-associate learning task.
\newblock Psychon Bull Rev 2020 Aug;27(4):791--800.

\bibitem[{Ashby and Maddox(2005)Ashby, F Gregory and Maddox, W
  Todd}]{Ashby2005-es}
Ashby FG, Maddox WT.
\newblock Human category learning.
\newblock Annu Rev Psychol 2005;56:149--178.

\bibitem[{Love et~al.(2004)Love, Bradley C and Medin, Douglas L and Gureckis,
  Todd M}]{Love2004-jl}
Love BC, Medin DL, Gureckis TM.
\newblock {SUSTAIN}: a network model of category learning.
\newblock Psychol Rev 2004 Apr;111(2):309--332.

\bibitem[{Zeithamova et~al.(2008)Zeithamova, Dagmar and Maddox, W Todd and
  Schnyer, David M}]{Zeithamova2008-rk}
Zeithamova D, Maddox WT, Schnyer DM.
\newblock Dissociable prototype learning systems: evidence from brain imaging
  and behavior.
\newblock J Neurosci 2008 Dec;28(49):13194--13201.

\bibitem[{Aizenstein et~al.(2000)Aizenstein, H J and MacDonald, A W and
  Stenger, V A and Nebes, R D and Larson, J K and Ursu, S and Carter, C
  S}]{Aizenstein2000-nn}
Aizenstein HJ, MacDonald AW, Stenger VA, Nebes RD, Larson JK, Ursu S, et~al.
\newblock Complementary category learning systems identified using
  event-related functional {MRI}.
\newblock J Cogn Neurosci 2000 Nov;12(6):977--987.

\bibitem[{Bowman and Zeithamova(2020)Bowman, Caitlin R and Zeithamova,
  Dagmar}]{Bowman2020-bp}
Bowman CR, Zeithamova D.
\newblock Training set coherence and set size effects on concept generalization
  and recognition.
\newblock J Exp Psychol Learn Mem Cogn 2020 Aug;46(8):1442--1464.

\bibitem[{Barsalou(1983)Barsalou, L W}]{Barsalou1983-ca}
Barsalou LW.
\newblock Ad hoc categories.
\newblock Mem Cognit 1983 May;11(3):211--227.

\bibitem[{Markman and Stilwell(2001)Markman, Arthur B and Stilwell, C
  Hunt}]{Markman2001-tn}
Markman AB, Stilwell CH.
\newblock Role-governed categories.
\newblock J Exp Theor Artif Intell 2001 Oct;13(4):329--358.

\bibitem[{Jones and Love(2007)Jones, Matt and Love, Bradley C}]{Jones2007-wx}
Jones M, Love BC.
\newblock Beyond common features: the role of roles in determining similarity.
\newblock Cogn Psychol 2007 Nov;55(3):196--231.

\bibitem[{Hoshi and Tanji(2000)Hoshi, E and Tanji, J}]{Hoshi2000-xl}
Hoshi E, Tanji J.
\newblock Integration of target and body-part information in the premotor
  cortex when planning action.
\newblock Nature 2000 Nov;408(6811):466--470.

\bibitem[{Wimmer and Shohamy(2012)Wimmer, G Elliott and Shohamy,
  Daphna}]{Wimmer2012-te}
Wimmer GE, Shohamy D.
\newblock Preference by association: how memory mechanisms in the hippocampus
  bias decisions.
\newblock Science 2012 Oct;338(6104):270--273.

\bibitem[{Mack et~al.(2013)Mack, Michael L and Preston, Alison R and Love,
  Bradley C}]{Mack2013-mj}
Mack ML, Preston AR, Love BC.
\newblock Decoding the brain's algorithm for categorization from its neural
  implementation.
\newblock Curr Biol 2013 Oct;23(20):2023--2027.

\bibitem[{Mok and Love(2020)Mok, Robert M and Love, Bradley C}]{Mok2020-pv}
Mok RM, Love BC.
\newblock Abstract Neural Representations of Category Membership beyond
  Information Coding Stimulus or Response.
\newblock J Cogn Neurosci 2020 Nov;p. 1--17.

\bibitem[{Tomlinson and Love(2006)Tomlinson, Marc T and Love, Bradley
  C}]{Tomlinson2006-fr}
Tomlinson MT, Love BC.
\newblock From Pigeons to Humans: Grounding Relational Learning in Concrete
  Examples.
\newblock In: Twentieth {AAAI}; 2006. p. 199--204.

\bibitem[{Zhou and Freedman(2019)Zhou, Yang and Freedman, David
  J}]{Zhou2019-eh}
Zhou Y, Freedman DJ.
\newblock Posterior parietal cortex plays a causal role in perceptual and
  categorical decisions.
\newblock Science 2019 Jul;365(6449):180--185.

\bibitem[{Zeithamova and Preston(2017)Zeithamova, Dagmar and Preston, Alison
  R}]{Zeithamova2017-ut}
Zeithamova D, Preston AR.
\newblock Temporal Proximity Promotes Integration of Overlapping Events.
\newblock J Cogn Neurosci 2017 Aug;29(8):1311--1323.

\bibitem[{Smith and Minda(1998)Smith, J David and Minda, John
  Paul}]{Smith1998-fe}
Smith JD, Minda JP.
\newblock Prototypes in the mist: The early epochs of category learning.
\newblock J Exp Psychol Learn Mem Cogn 1998 Nov;24(6):1411--1436.

\bibitem[{Kruschke(1992)Kruschke, J K}]{Kruschke1992-yj}
Kruschke JK.
\newblock {ALCOVE}: an exemplar-based connectionist model of category learning.
\newblock Psychol Rev 1992 Jan;99(1):22--44.

\bibitem[{Medin and Schaffer(1978)Medin, Douglas L and Schaffer, Marguerite
  M}]{Medin1978-ex}
Medin DL, Schaffer MM.
\newblock Context theory of classification learning.
\newblock Psychol Rev 1978;85(3):207--238.

\bibitem[{Shepard(1957)Shepard, Roger N}]{Shepard1957-xl}
Shepard RN.
\newblock Stimulus and response generalization: A stochastic model relating
  generalization to distance in psychological space.
\newblock Psychometrika 1957 Dec;22(4):325--345.

\bibitem[{Poppenk et~al.(2013)Poppenk, Jordan and Evensmoen, Hallvard R and
  Moscovitch, Morris and Nadel, Lynn}]{Poppenk2013-uy}
Poppenk J, Evensmoen HR, Moscovitch M, Nadel L.
\newblock Long-axis specialization of the human hippocampus.
\newblock Trends Cogn Sci 2013 May;17(5):230--240.

\bibitem[{Brunec et~al.(2018)Brunec, Iva K and Bellana, Buddhika and Ozubko,
  Jason D and Man, Vincent and Robin, Jessica and Liu, Zhong-Xu and Grady,
  Cheryl and Rosenbaum, R Shayna and Winocur, Gordon and Barense, Morgan D and
  Moscovitch, Morris}]{Brunec2018-uk}
Brunec IK, Bellana B, Ozubko JD, Man V, Robin J, Liu ZX, et~al.
\newblock Multiple Scales of Representation along the Hippocampal
  Anteroposterior Axis in Humans.
\newblock Curr Biol 2018 Jul;28(13):2129--2135.e6.

\bibitem[{Collin et~al.(2015)Collin, Silvy H P and Milivojevic, Branka and
  Doeller, Christian F}]{Collin2015-ky}
Collin SHP, Milivojevic B, Doeller CF.
\newblock Memory hierarchies map onto the hippocampal long axis in humans.
\newblock Nat Neurosci 2015 Nov;18(11):1562--1564.

\bibitem[{Kjelstrup et~al.(2008)Kjelstrup, Kirsten Brun and Solstad, Trygve and
  Brun, Vegard Heimly and Hafting, Torkel and Leutgeb, Stefan and Witter, Menno
  P and Moser, Edvard I and Moser, May-Britt}]{Kjelstrup2008-he}
Kjelstrup KB, Solstad T, Brun VH, Hafting T, Leutgeb S, Witter MP, et~al.
\newblock Finite scale of spatial representation in the hippocampus.
\newblock Science 2008 Jul;321(5885):140--143.

\bibitem[{Cohn-Sheehy et~al.(2021)Cohn-Sheehy, Brendan I and Delarazan,
  Angelique I and Crivelli-Decker, Jordan E and Reagh, Zachariah M and Mundada,
  Nidhi S and Yonelinas, Andrew P and Zacks, Jeffrey M and Ranganath,
  Charan}]{Cohn-Sheehy2021-cm}
Cohn-Sheehy BI, Delarazan AI, Crivelli-Decker JE, Reagh ZM, Mundada NS,
  Yonelinas AP, et~al.
\newblock Narratives bridge the divide between distant events in episodic
  memory.
\newblock Mem Cognit 2021 Apr;.

\bibitem[{Cai et~al.(2016)Cai, Denise J and Aharoni, Daniel and Shuman, Tristan
  and Shobe, Justin and Biane, Jeremy and Song, Weilin and Wei, Brandon and
  Veshkini, Michael and La-Vu, Mimi and Lou, Jerry and Flores, Sergio E and
  Kim, Isaac and Sano, Yoshitake and Zhou, Miou and Baumgaertel, Karsten and
  Lavi, Ayal and Kamata, Masakazu and Tuszynski, Mark and Mayford, Mark and
  Golshani, Peyman and Silva, Alcino J}]{Cai2016-nt}
Cai DJ, Aharoni D, Shuman T, Shobe J, Biane J, Song W, et~al.
\newblock A shared neural ensemble links distinct contextual memories encoded
  close in time.
\newblock Nature 2016 Jun;534(7605):115--118.

\bibitem[{Rashid et~al.(2016)Rashid, Asim J and Yan, Chen and Mercaldo,
  Valentina and Hsiang, Hwa-Lin Liz and Park, Sungmo and Cole, Christina J and
  De Cristofaro, Antonietta and Yu, Julia and Ramakrishnan, Charu and Lee, Soo
  Yeun and Deisseroth, Karl and Frankland, Paul W and Josselyn, Sheena
  A}]{Rashid2016-qa}
Rashid AJ, Yan C, Mercaldo V, Hsiang HLL, Park S, Cole CJ, et~al.
\newblock Competition between engrams influences fear memory formation and
  recall.
\newblock Science 2016 Jul;353(6297):383--387.

\bibitem[{Silva et~al.(2009)Silva, Alcino J and Zhou, Yu and Rogerson, Thomas
  and Shobe, Justin and Balaji, J}]{Silva2009-eq}
Silva AJ, Zhou Y, Rogerson T, Shobe J, Balaji J.
\newblock Molecular and cellular approaches to memory allocation in neural
  circuits.
\newblock Science 2009 Oct;326(5951):391--395.

\bibitem[{Ezzyat and Davachi(2014)Ezzyat, Youssef and Davachi,
  Lila}]{Ezzyat2014-vc}
Ezzyat Y, Davachi L.
\newblock Similarity breeds proximity: pattern similarity within and across
  contexts is related to later mnemonic judgments of temporal proximity.
\newblock Neuron 2014 Mar;81(5):1179--1189.

\bibitem[{Hsieh et~al.(2014)Hsieh, Liang-Tien and Gruber, Matthias J and
  Jenkins, Lucas J and Ranganath, Charan}]{Hsieh2014-go}
Hsieh LT, Gruber MJ, Jenkins LJ, Ranganath C.
\newblock Hippocampal activity patterns carry information about objects in
  temporal context.
\newblock Neuron 2014 Mar;81(5):1165--1178.

\bibitem[{Sanchez and Zeithamova(2020)Sanchez, Maria Alexandra de Araujo and
  Zeithamova, Dagmar}]{Sanchez2020-jw}
Sanchez MAdA, Zeithamova D.
\newblock Generalization and source memory in acquired equivalence; 2020, -.

\bibitem[{Keresztes et~al.(2018)Keresztes, Attila and Ngo, Chi T and
  Lindenberger, Ulman and Werkle-Bergner, Markus and Newcombe, Nora
  S}]{Keresztes2018-pl}
Keresztes A, Ngo CT, Lindenberger U, Werkle-Bergner M, Newcombe NS.
\newblock Hippocampal Maturation Drives Memory from Generalization to
  Specificity.
\newblock Trends Cogn Sci 2018 Aug;22(8):676--686.

\bibitem[{Johansen and Palmeri(2002)Johansen, Mark K and Palmeri, Thomas
  J}]{Johansen2002-hu}
Johansen MK, Palmeri TJ.
\newblock Are there representational shifts during category learning?
\newblock Cogn Psychol 2002 Dec;45(4):482--553.

\bibitem[{Bright and Feeney(2014)Bright, Aim{\'e}e K and Feeney,
  Aidan}]{Bright2014-hs}
Bright AK, Feeney A.
\newblock The engine of thought is a hybrid: roles of associative and
  structured knowledge in reasoning.
\newblock J Exp Psychol Gen 2014 Dec;143(6):2082--2102.

\bibitem[{Wills et~al.(2011)Wills, Andy J and Graham, Steven and Koh, Zhisheng
  and McLaren, Ian P L and Rolland, Matthew D}]{Wills2011-pb}
Wills AJ, Graham S, Koh Z, McLaren IPL, Rolland MD.
\newblock Effects of concurrent load on feature- and rule-based generalization
  in human contingency learning.
\newblock J Exp Psychol Anim Behav Process 2011 Jul;37(3):308--316.

\bibitem[{Pace-Schott et~al.(2009)Pace-Schott, Edward F and Milad, Mohammed R
  and Orr, Scott P and Rauch, Scott L and Stickgold, Robert and Pitman, Roger
  K}]{Pace-Schott2009-qi}
Pace-Schott EF, Milad MR, Orr SP, Rauch SL, Stickgold R, Pitman RK.
\newblock Sleep promotes generalization of extinction of conditioned fear.
\newblock Sleep 2009 Jan;32(1):19--26.

\bibitem[{Lau et~al.(2010)Lau, H and Tucker, M A and Fishbein, W}]{Lau2010-eu}
Lau H, Tucker MA, Fishbein W.
\newblock Daytime napping: Effects on human direct associative and relational
  memory.
\newblock Neurobiol Learn Mem 2010 May;93(4):554--560.

\bibitem[{Schapiro et~al.(2018)Schapiro, Anna C and McDevitt, Elizabeth A and
  Rogers, Timothy T and Mednick, Sara C and Norman, Kenneth
  A}]{Schapiro2018-ig}
Schapiro AC, McDevitt EA, Rogers TT, Mednick SC, Norman KA.
\newblock Human hippocampal replay during rest prioritizes weakly learned
  information and predicts memory performance.
\newblock Nat Commun 2018 Sep;9(1):3920.

\bibitem[{Maddox et~al.(2009)Maddox, W Todd and Glass, Brian D and Wolosin,
  Sasha M and Savarie, Zachary R and Bowen, Christopher and Matthews, Michael D
  and Schnyer, David M}]{Maddox2009-fj}
Maddox WT, Glass BD, Wolosin SM, Savarie ZR, Bowen C, Matthews MD, et~al.
\newblock The effects of sleep deprivation on information-integration
  categorization performance.
\newblock Sleep 2009 Nov;32(11):1439--1448.

\bibitem[{Lerner and Gluck(2019)Lerner, Itamar and Gluck, Mark
  A}]{Lerner2019-fj}
Lerner I, Gluck MA.
\newblock Sleep and the extraction of hidden regularities: A systematic review
  and the importance of temporal rules.
\newblock Sleep Med Rev 2019 Oct;47:39--50.

\bibitem[{Poldrack and Foerde(2008)Poldrack, Russell A and Foerde,
  Karin}]{Poldrack2008-de}
Poldrack RA, Foerde K.
\newblock Category learning and the memory systems debate.
\newblock Neurosci Biobehav Rev 2008;32(2):197--205.

\bibitem[{Sakagami and Tsutsui(1999)Sakagami, M and Tsutsui,
  K}]{Sakagami1999-rs}
Sakagami M, Tsutsui K.
\newblock The hierarchical organization of decision making in the primate
  prefrontal cortex.
\newblock Neurosci Res 1999 Jul;34(2):79--89.

\bibitem[{Shima et~al.(2007)Shima, Keisetsu and Isoda, Masaki and Mushiake,
  Hajime and Tanji, Jun}]{Shima2007-qu}
Shima K, Isoda M, Mushiake H, Tanji J.
\newblock Categorization of behavioural sequences in the prefrontal cortex.
\newblock Nature 2007 Jan;445(7125):315--318.

\bibitem[{Tomita et~al.(1999)Tomita, H and Ohbayashi, M and Nakahara, K and
  Hasegawa, I and Miyashita, Y}]{Tomita1999-ki}
Tomita H, Ohbayashi M, Nakahara K, Hasegawa I, Miyashita Y.
\newblock Top-down signal from prefrontal cortex in executive control of memory
  retrieval.
\newblock Nature 1999 Oct;401(6754):699--703.

\end{thebibliography}

\end{document}